\documentclass[prd,nofootinbib,showpacs,preprint]{revtex4}
\usepackage{graphicx}
\usepackage{epsfig}
\usepackage{amsmath}
\usepackage{amsfonts}
\usepackage{amssymb}
\usepackage{url}
\usepackage{hyperref}
\usepackage{subfigure}
\usepackage{color,graphicx}
\usepackage{cancel}
\newcommand{\beqa}{\begin{eqnarray}}
\newcommand{\eeqa}{\end{eqnarray}}
\newcommand{\beq}{\begin{equation}}
\newcommand{\eeq}{\end{equation}}

\newcommand{\nn}{\nonumber}
\newcommand{\bmt}{\begin{pmatrix}}
\newcommand{\emt}{\end{pmatrix}}
\usepackage[toc,page]{appendix}
\usepackage{comment}
\newcommand{\be}{\begin{equation}}
\newcommand{\ee}{\end{equation}}
\newcommand{\bea}{\begin{eqnarray}}
\newcommand{\eea}{\end{eqnarray}}

\begin{document}
\title{Impact of vector leptoquark on $\bar B \to \bar K^* l^+ l^-$ anomalies }
\author{Suchismita Sahoo and Rukmani Mohanta }
\affiliation{\,School of Physics, University of Hyderabad,
              Hyderabad - 500046, India  }
\begin{abstract}
Motivated by the recent measurement of the lepton flavour nonuniversality ratio $R_{K^*}$ by the LHCb Collaboration,
we study the implications of vector leptoquarks on the observed anomalies associated with  the $\bar B \to \bar K^* l^+ l^-$ decay processes. The leptoquark couplings are constrained from the measured  branching ratios of    $B_s \to l^+ l^-$, $K_L \to l^+ l^-$ and $B_s \to \mu^\mp e^\pm$ processes.  Using these constrained couplings, we  estimate the branching ratios, forward-backward and lepton polarization asymmetries and also the form factor independent  optimized observables ($P_{i}^{(\prime)}$)  for $\bar B \to \bar K^* l^+ l^-$ modes in the high recoil limit.   We   also study the  other lepton flavour universality violating observables, such as $Q_{F_{L, T}}$,   $Q_{i}$ and $B_{5, 6s}$, where $i=1,2,\cdots6, 8$.   Furthermore, we investigate the lepton flavour violating $K_L \to \mu^\mp e^\pm$ decay process in this model.

\end{abstract}
\maketitle
\section{Introduction}

In recent times $B$ physics is going through a challenging phase,  several anomalies at the level of  $(3-4) \sigma$ \cite{RK-exp, RDstar-LHCb, RKstar-exp, phi-decayrate, Kstar-decayrate, P5p, isospin-kstar} have been observed by the LHCb Collaboration in the rare flavour changing neutral current (FCNC) processes involving the quark level transition $b \to s l^+ l^-$. 
As these processes are one-loop suppressed in the standard model (SM), they may  play  a vital role to  decipher the signature of new physics (NP) beyond it. To supplement these observations, 
recently  LHCb has reported $2.2 \sigma$ and $2.4\sigma$ discrepancies in the measurement of  $R_{K^*}$ observable in the dilepton invariant mass squared bins $q^2 \in [0.045, 1.1]~{\rm GeV}^2$ and $q^2 \in [1.1, 6.0]~{\rm GeV}^2$ \cite{RKstar-exp}, which are in the same line as the previous result on the violation of lepton universality parameter $R_K$ \cite{RK-exp}.  Also the  lepton nonuniversality (LNU) parameters in the $B \to D^{(*)}$ processes ($R_{D^{(*)}}$) have been measured by   Belle, BaBar and LHCb collaborations,    which have  respectively $1.9 \sigma$ \cite{RD-BaBar, RD-exp} and $3.3\sigma$ \cite{RDstar-LHCb, RD-exp} deviations from their corresponding SM predictions.   In Table I, we present the observed   LNU ratios,  associated with the $b \to s l^+ l^-$ and $b \to c l \nu_l$ processes at  LHCb and $B$ factories. Furthermore, the decay rate of $B_s \to \phi \mu^+ \mu^-$ process \cite{phi-decayrate} also has discrepancy of around $3\sigma$ in the low $q^2$ region.

\begin{table}[htb]
\begin{center}
\caption{The LNU parameters observed by the LHCb collaboration  and the  $B$ factories.}
\vspace*{0.1 true in}
\begin{tabular}{|c|c|c| c | c|}
\hline
LNU parameters  ~ & ~SM predictions ~&~ Expt. result  & ~ Deviation~ \\

\hline

$R_K|_{q^2 \in [1.0, 6.0]}$  ~&  ~ $1.0003 \pm 0.0001$~\cite{RK-SM} ~&~ $0.745^{+0.090}_{-0.074} \pm 0.036$~\cite{RK-exp}~& ~$2.6 \sigma$~\\

\hline

$R_{K^*}|_{q^2 \in [0.045, 1.1]} $ ~& ~ $0.92 \pm 0.02$ ~\cite{RKstar-SM}~& ~$0.66^{+0.11}_{-0.07} \pm 0.03$~\cite{RKstar-exp}~&~ $2.2 \sigma$~\\

\hline

$R_{K^*}|_{q^2 \in [1.1, 6.0]} $ ~& ~$ 1.00 \pm 0.01 $~\cite{RKstar-SM}~& $0.69^{+0.11}_{-0.07} \pm 0.05$ ~\cite{RKstar-exp}~&~ $2.4 \sigma$~\\

\hline

$R_D$ ~& ~$ 0.300 \pm  0.008 $~\cite{RD-SM}~&~$0.397 \pm 0.040 \pm 0.028$ ~\cite{RD-exp}~& ~$1.9 \sigma$~\\

\hline

$R_{D^*} $ ~&~ $0.252 \pm  0.003$ ~\cite{RDstar-SM}~& ~$0.316 \pm 0.016 \pm 0.010 $~\cite{RD-exp}~&~$3.3 \sigma$~ \\

\hline
\end{tabular}
\end{center}
\end{table}

In this context, we would like to investigate whether  the observed anomalies in the rare $\bar B \to \bar K^* l^+ l^-$ decay processes, mediated through $b \to s l^+ l^-$ transitions, can be explained in the vector leptoquark model. 
In the last few years, these processes  have  provided several surprising   results   and  played a very crucial role to look for NP signals, as the measurement of four-body angular distribution provides a large number of observables which can be used to probe NP signature.    In the low $q^2 \in [1, 6]~{\rm GeV}^2 $ region (where $q^2$ denotes the dilepton invariant mass square), the theoretical  predictions for such observables are very  precise and generally free from the hadronic uncertainties. However, the observed forward-backward asymmetry is systematically below the corresponding  SM prediction, though the zero crossing point is consistent with it.
Moreover, the LHCb Collaboration has reported  many other deviations from the SM expectations  in the angular observables. The largest discrepancy of $\sim3 \sigma$   in   the famous $P_5^\prime$ optimized observable \cite{P5p} and the decay rate \cite{Kstar-decayrate}  of these processes provide a  sensitive probe to explore NP effects in $b \to s \gamma$, $b \to s ll$ transitions.  In addition, the isospin asymmetry \cite{isospin-kstar} is also measured by the LHCb experiment in the full $q^2$ region, which can be used to probe the  NP signal. For the first time, recently Belle has measured two new lepton flavour universality  violating (LFUV) observables $Q_{4, 5} = P_{4, 5}^{\mu^\prime} - P_{4, 5}^{e^\prime} $ \cite{Q4-Belle}.   In order to scrutinize the above results, these processes have already been investigated  in the context of    various new physics models and also in   model-independent ways. The recent measurement on the $R_{K^*}$  parameter at LHCb experiment has drawn much attention  to restudy these processes in the low $q^2$ region. In the light of recent $R_{K^*}$ data,   several works
\cite{recent-arXiv}, have been reported in the literature recently.

To understand the origin of the  current issues observed at LHCb experiment in a particular theoretical framework,  here we extend the SM by adding a single vector leptoquark (LQ) and  reinvestigate the rare semileptonic  $\bar B \to \bar K^* l^+ l^-$ decay processes.   Though there are a few recent studies in  the literature \cite{recent-arXiv}, which  have investigated $R_{K^*}$ anomalies but  no analysis of $R_{K^*}$ has been done  with vector LQ, which can induce the process at tree level. In our previous work \cite{mohanta0}, we have made a comparative study of the rare semileptonic  $\bar B \to \bar K^* l^+ l^-$ decay modes in both the $(3,2,7/6)$ and $(3,2,1/6)$ scalar LQ models. However, we have not investigated the new  $R_{K^*}$, $Q_{4,5}$ and  $Q_{F_{L,T}}$ observables.   The model-independent analysis of these new set of observables can be found in Ref. \cite{kstar-Q4}.  The motivation of this work is to check how the angular analysis of $\bar B \to \bar K^* l^+ l^-$ processes in the context of vector  LQ could help establishing the  possible existence of NP from the above discussed anomalies.  LQs  are hypothetical  color triplet bosonic particles, which arise naturally from the unification of quarks and leptons, and carry both the lepton and baryon numbers. They can be either scalar (spin 0) or vector (spin 1) in nature. The presence of  vector LQs at the TeV scale can be found in many extended SM theories such as grand unified theories based on $SU(5)$, $SO(10)$, etc. \cite{GUT, Pati}, Pati-Salam model \cite{Pati, Pati-salam}, composite model \cite{Composite} and the technicolor model \cite{Technicolor}. The baryon number conserving LQs avoid proton decay and  could be light enough to be seen in the current experiments. Thus, in this work,  we consider  the singlet  $(3, 1, 2/3)$ vector LQ, which is invariant under the SM $SU(3)_C \times SU(2)_L \times U(1)_Y$ gauge group  and conserves both the baryon and lepton numbers. In addition to the (axial)vector operators, this LQ also provides additional (pseudo)scalar operators to the SM.  We compute the branching ratios, forward-backward asymmetries, polarization asymmetries and the form factor independent (FFI)  observables $(P_i^{(\prime)})$ of  the  $\bar B \to \bar  K^* l^+ l^-$ processes in this  model. 
In this paper,  we  mainly focus on  the $R_{K^*}$ anomaly and the additional observables related to  the lepton flavour violation  in  order to confirm or rule out the presence of lepton nonuniversality in the rare  $B$ meson decays.  We also investigate the $Q_i$, $Q_{F_{L,T}}$ and $B_i$ observables in the context of vector LQ, so as to reveal the possible interplay of NP.  In the literature, the observed  anomalies at LHCb experiment  in various rare decays of $B$ mesons  have been studied  in the LQ model  \cite{mohanta0, mohanta1, mohanta2, mohanta3, davidson, kosnik-LQ, KL-LQ}.

The paper is organized as follows.  In section II, we discuss the effective Hamiltonian responsible for the $b \to s l^+ l^-$ processes and the new physics contributions arising due to the exchange of vector LQ. In section III, we show the constraints on LQ couplings from the branching ratios of  rare $B_s \to l^+ l^-$, $K_L \to l^+ l^-$ and $B_s \to \mu^\mp e^\pm$ processes.  The branching ratios, forward-backward asymmetries,  lepton polarizations and the  CP violating parameters in the  $\bar B \to \bar  K^* l^+ l^-$ processes are calculated in section IV. Section V deals with the lepton flavour violating decay $K_L \to \mu^{\mp} e^{\pm}$ and section VI   contains the summary.

\section{Generalized effective Hamiltonian}
In the SM, the most general effective Hamiltonian responsible for the quark level transitions $b \to s l^+ l^-$ is given by
\cite{b-s-Hamiltonian} 
\bea
{\cal H}_{\rm eff} &=& - \frac{ 4 G_F}{\sqrt 2} V_{tb} V_{ts}^* \Bigg[\sum_{i=1}^6 C_i(\mu) \mathcal{O}_i +C_7 \frac{e}{16 \pi^2} \Big(\bar s \sigma_{\mu \nu}
(m_s P_L + m_b P_R ) b\Big) F^{\mu \nu} \nn\\
&&+ \frac{\alpha}{4 \pi}\left (C_9^{\rm eff} (\bar s \gamma^\mu P_L b) (\bar l \gamma_\mu l )+ C_{10} (\bar s \gamma^\mu P_L b)
(\bar l \gamma_\mu \gamma_5 l)\right )\Bigg]\;,\label{ham}
\eea
where $G_F$ is the Fermi constant, $V_{qq^\prime}$ are the Cabibbo-Kobayashi-Maskawa  (CKM) matrix elements, $\alpha$ is the fine structure constant and $P_{L, R} =(1\mp \gamma_5)/2$ are the  chiral  projection operators. Here $\mathcal{O}_i$'s are the six dimensional operators and  $C_i$'s are the corresponding Wilson coefficients,   evaluated at the renormalization  scale $\mu =m_b$ \cite{b-s-Wilson}.

\subsection{Contributions from vector leptoquark }

The SM effective Hamiltonian (\ref{ham}) can be modified by adding a single vector LQ and  will give measurable deviations from the corresponding SM predictions in the $B$ sector.  Here we consider $V^{1}(3, 1, 2/3)$ singlet vector LQ which is invariant  under the SM gauge group $SU(3)_C \times SU(2)_L \times U(1)_Y$.  In order to avoid rapid proton decay, we assume that the LQ conserves both baryon and lepton numbers.  The baryon  number conserving vector LQs  can have sizeable Yukawa couplings and could be light enough to be accessible in a current collider. The $V^{1}(3, 1, 2/3)$ LQ could potentially contribute to the $b \to s l^+ l^- $ processes and one can constrain the 
 corresponding LQ couplings from the experimental data on $B_s \to l^+ l^-$ processes.  

The interaction Lagrangian for $V^{1}(3,1,2/3)$ leptoquark    is given by \cite{kosnik-LQ, mohanta3}
\begin{equation}
  \mathcal{L}^{(1)} = \left(g_L\, \overline{Q}_L \gamma^\mu L_L  +  g_R\, \overline{d_R}
    \gamma^\mu l_R \right)\, V^{1}_\mu + {\rm h.c.},
\end{equation}
where $Q_L~(L_L)$ is the left-handed quark (lepton) doublets and  $d_R~(l_R)$ is the right-handed down-type quark (charged-lepton) singlets. Here $g_L$ is the coupling of vector LQ with the quark and lepton doublets and $g_R$ is the LQ coupling with down-type quarks and the right-handed leptons.  To keep the notations clean, the leptoquark couplings $g_L$ and $g_R$  are considered in the mass basis of down-type quarks, i.e.,  the couplings $g_L$ and $g_R$ are rotated and expressed in the quark mass basis by the redefinition $U_L^\dagger g_L \rightarrow g_L$ and $U_R^\dagger g_R \rightarrow g_R$, where $U_{L,R}$ connect the mass and gauge bases, i.e., $d_{L,R}^{\rm gauge}=U_{L,R} d_{L,R}^{\rm mass}$.  The interaction Lagrangian (2) provides in addition to the vector ($C_{9}^{(\prime)\rm LQ }$) and axial-vector ($C_{10}^{(\prime)\rm LQ}$) new Wilson coefficients, new scalar $C_{S}^{(\prime)\rm LQ}$ and pseudoscalar $C_{P}^{(\prime)\rm LQ}$ coefficients, and is thus non-chiral in nature. The new Wilson coefficients  are related to the LQ couplings through the following  relations \cite{kosnik-LQ, mohanta3}
\begin{subequations}
\bea
 && C_9^{\rm LQ} = -C_{10}^{\rm LQ} = \frac{\pi}{\sqrt{2} G_F V_{tb}V_{ts}^* \alpha}
  \frac{(g_L)_{s l} (g_L)^*_{bl}}{M_{\rm LQ}^2}\,,\label{u1c10np}\\
&&  C_9^{\prime \rm LQ} = C_{10}^{\prime \rm LQ} = \frac{\pi}{\sqrt{2} G_F V_{tb}V_{ts}^*
    \alpha} \frac{(g_R)_{sl} (g_R)^*_{bl}}{M_{\rm LQ}^2}\,,\label{u1c10pnp} \\ 
&&  -C_P^{\rm LQ} = C_{S}^{\rm LQ} = \frac{\sqrt{2}\pi}{G_F V_{tb}V_{ts}^* \alpha}
  \frac{(g_L)_{sl} (g_R)^*_{bl}}{M_{\rm LQ}^2}\,,\label{u1csnp} \\
&&  C_P^{\prime \rm LQ} = C_{S}^{\prime  \rm LQ}= \frac{\sqrt{2}\pi}{G_F V_{tb}V_{ts}^* \alpha} \frac{(g_R)_{sl} (g_L)^*_{bl}}{M_{\rm LQ}^2}\, \label{u1cspnp}. 
\eea
\end{subequations}

\section{Constraint on the vector leptoquark couplings}
After knowing about the  interplay of possible new Wilson coefficients, we  now proceed to constrain the new physics parameters by comparing the theoretical and experimental results on various rare $B(K)$ meson decays. 

\subsection{$B_s \to l^+ l^-$ processes}
In this subsection, we show the constraints on the new  LQ couplings from the $B_s \to l^+ l^-$ processes, as these new coefficients also  contribute to the $B_s \to l^+ l^-$ processes. These decay processes are very rare in the SM as they  occur at loop level and  further suffer from helicity suppression. The only
non-perturbative quantity involved  is the  decay constant of $B$ mesons, which can be reliably
calculated  by using the non-perturbative methods, thus, these processes are theoretically very clean. In the SM, only the $C_{10}^{\rm SM}$ Wilson coefficient contributes to the branching ratio. 

The branching ratios of $B_s \to l^+ l^-$ processes in the  vector LQ model are given by \cite{Buras}
\bea
{\rm BR}(B_s \to l^+ l^-) = \frac{G_F^2}{16 \pi^3} \tau_{B_s} \alpha^2 f_{B_s}^2 M_{B_s} m_{l}^2 |V_{tb} V_{ts}^*|^2
\left |C_{10}^{\rm SM}\right |^2 \sqrt{1- \frac{4 m_l^2}{M_{B_s}^2}} \times \left(|P|^2 + |S|^2 \right),
\eea
where $P$ and $S$ parameters are defined as
\bea
&&P \equiv \frac{C_{10}^{\rm SM}+ C_{10}^{\rm LQ}-C_{10}^{\prime \rm LQ}}{C_{10}^{\rm SM}}+\frac{M_{B_s}^2}{2m_{l}} \Big(\frac{m_b}{m_b+m_s} \Big) \Big(\frac{C_{P}^{\rm LQ}-C_{P}^{\prime \rm LQ}}{C_{10}^{\rm SM}}\Big),\nn \\ 
&&S \equiv \sqrt{1- \frac{4 m_l^2}{M_{B_s}^2}} \frac{M_{B_s}^2}{2m_{l}} \Big(\frac{m_b}{m_b+m_s} \Big) \Big(\frac{C_{S}^{\rm LQ}-C_{S}^{\prime \rm LQ}}{C_{10}^{\rm SM}}\Big). 
\label{P-S}
\eea
Now to compare the theoretical branching ratios with the experimental results, one can define the parameter $R_q$, which is the ratio of branching fraction to its SM value  as 
\bea
R_{q}=\frac{{\rm BR}(B_s \to l^+ l^-)}{{\rm BR}^{\rm SM}(B_s \to l^+ l^-)} = |P|^2+|S|^2.
\label{R-q}
\eea
Using Eqn. (\ref{R-q}), 
we constrain the new couplings by comparing the SM predicted branching ratios \cite{Bobeth} of $B_s \to l^+ l^-$ processes with their corresponding experimental results \cite{e-br, mu-br, tau-br}.
 The constraint on vector LQ couplings from $B_s \to l^+ l^-$ processes has already been extracted in \cite{mohanta1, mohanta3}, therefore, here we will simply quote the results. In Table II, we have presented the obtained  bound on the $(g_L)_{sl}(g_L)^*_{b l}$ leptoquark couplings. The constraints on the combination of $C_S^{(\prime) \rm  LQ}$ Wilson coefficients i.e., $C_S^{\rm LQ} \pm C_S^{\prime \rm LQ}$ are presented in Table III, from which one can obtain the bound on individual $C_S^{(\prime) \rm  LQ}$ Wilson coefficients.

\subsection{$K_L \to l^+ l^-$ processes}
The effective Hamiltonian responsible for the  $s \to d l^+ l^-$ quark level transitions in the SM  is given by \cite{KL-2}
\bea 
\mathcal{H}_{\rm eff} &=& \frac{G_F}{\sqrt{2}} \frac{\alpha}{2 \pi \sin^2\theta_W} \left(\lambda_c Y_{NL} + \lambda_t Y(x_t) \right)  \left( \bar{s} \gamma^\mu (1-\gamma_5)d \right) \left(\bar{l} \gamma_\mu (1-\gamma_5)l \right)\\
&=& \frac{G_F}{\sqrt{2}} \frac{\alpha}{2 \pi} \lambda_u C^{\rm SM}_K \left( \bar{s} \gamma^\mu (1-\gamma_5)d \right) \left(\bar{l} \gamma_\mu (1-\gamma_5)l \right),
\label{sd-ham}
\eea
where $\lambda_i = V_{id} V_{is}^*$,  $x_t = m_t^2/M_W^2$,  $\sin^2 \theta_W = 0.23$ and $C^{\rm SM}_K$ is the  SM Wilson coefficient  given as
\bea
C^{\rm SM}_K = \frac{\left(\lambda_c Y_{NL} + \lambda_t Y(x_t) \right)}{\sin^2\theta_W \lambda_u}\;.
\eea
Here the functions $Y_{NL}$ and $Y(x_t)$ \cite{KL-3} are the contributions from the charm and top quark respectively. The estimated branching ratio of the short distance (SD) part of the  $K_L \to \mu^+ \mu^-$ process is ${\rm BR}(K_L \to \mu^+ \mu^-)|_{\rm SD} < 2.5 \times 10^{-9}$ \cite{KL-1}. 

Including  the   $(3,1,2/3)$ vector LQ contributions, the total branching ratios of $K_L \to l^+ l^-$ processes are  given by \cite{mohanta3}
\bea \label{br-KL}
{\rm BR}(K_L \to l^+ l^-) = \frac{G_F^2}{8 \pi^3} \tau_{K_L} \alpha ^2 f_{K}^2 M_{K} m_{l}^2 |\lambda_u|^2
\left |C_{\rm SM}^{K}\right |^2 \sqrt{1- \frac{4 m_l^2}{M_{K}^2}} \times \left(|P_K|^2 + |S_K|^2 \right),
\eea
where 
$P_K $ and $S_K $ parameters  have analogous expressions as Eqn. (\ref{P-S}) with the replacement   of $M_{B_s} \to M_K$ and  the corresponding new Wilson coefficients by $C_{iK}^{\rm LQ}$, which are given as \cite{mohanta3}
\begin{subequations}
\bea
&&C_{10K}^{\rm LQ} = -\frac{\pi}{G_F \alpha \lambda_u} \frac{{\rm Re}[(g_L)_{dl}(g_L)_{sl}^*]}{M_{\rm LQ}^2}\;, \\
&&C_{10K}^{\prime \rm  LQ} = -\frac{\pi}{G_F \alpha \lambda_u} \frac{{\rm Re}[(g_R)_{dl}(g_R)_{sl}^*]}{M_{\rm LQ}^2}\;,\\
&&C_{SK}^{\rm LQ} =-C_{PK}^{\rm LQ} = \frac{\pi}{2 G_F \alpha \lambda_u} \frac{{\rm Re}[(g_L)_{dl}(g_R)_{sl}^*]}{M_{\rm LQ}^2}\;, \\
&&C_{SK}^{\prime \rm LQ} =C_{PK}^{\prime \rm LQ} = \frac{\pi}{2 G_F \alpha  \lambda_u} \frac{{\rm Re}[(g_R)_{dl}(g_L)_{sl}^*]}{M_{\rm LQ}^2}\;.
\eea
\end{subequations}
Now using the experimental upper limits \cite{pdg} on the branching ratios of  $K_L \to l^+ l^-$ decay processes, the constraint on the new physics parameters are extracted  in Ref. \cite{mohanta3}. In Table II, we present the constraints on $(g_L)_{dl}(g_L)_{sl}^*$ couplings, and the bound on $C_S^{\rm LQ} \pm C_S^{\prime \rm LQ}$ Wilson coefficients are given in Table III. 
\begin{table}[htb]
\begin{center}
\caption{Constraints on the LQ couplings obtained from  various leptonic $B_{s} \to l^+ l^-$ and $K_L \to l^+ l^-$ decay processes .}
\vspace*{0.1 true in}
\begin{tabular}{|c|c|c|}
\hline
Decay Process ~& ~Couplings involved ~&~ Upper bound of  \\
             &  &~the couplings   \\
\hline

$B_s \to e^\pm e^\mp $ &~ $|(g_L)_{s e} (g_L)^*_{b e}|$ ~& ~$ <~ 11.8 $~\\

\hline

$B_s \to \mu^\pm \mu^\mp $ &~ $|(g_L)_{s \mu} (g_L)^*_{b \mu}|$ ~& ~$ \leq 2.3 \times 10^{-3} $~\\

\hline

$K_L \to e^\pm e^\mp $ &~ $|(g_L)_{d e} (g_L)^*_{s e}|$ ~& ~$ (1.3-2.35) \times 10^{-3} $~\\

\hline

$K_L \to \mu^\pm \mu^\mp $ &~ $|(g_L)_{d \mu} (g_L)^*_{s \mu}|$ ~& ~$(1.4-1.5) \times 10^{-4} $~\\

\hline
\end{tabular}
\end{center}
\end{table}


\begin{table}[htb]
\begin{center}
\caption{Constraint on  combinations of $C_{S(K)}^{(\prime)\rm LQ}$ Wilson coefficients from rare leptonic $B_s \to l^+ l^-$ and $K_L \to l^+ l^-$ decay processes.}
\vspace*{0.1 true in}
\begin{tabular}{|c|c|c|}
\hline
Decay Process ~& ~Bound on $C_{S(K)}^{\rm LQ}+ C_{S(K)}^{\prime \rm LQ}$   ~&~Bound on $C_{S(K)}^{\rm LQ}- C_{S(K)}^{\prime  \rm LQ}$  \\
\hline

$B_s \to e^\pm e^\mp $ &~ $-1.4 \to 1.4$ ~& ~$-1.4 \to 1.4$ ~\\

\hline
$B_s \to \mu^\pm \mu^\mp $~~ &~~ $0.0 \to 0.32$ ~~& ~~$0.1 \to 0.18$~\\

\hline

$K_L \to e^\pm e^\mp $ &~ $(-2.0 \to 2.0) \times 10^{-4}$ ~& ~$(1.25 \to 2) \times 10^{-4}$ ~\\

\hline
$K_L \to \mu^\pm \mu^\mp $~~ &~~ $(-6.0 \to 3.0) \times 10^{-3}$ ~~& ~~$(0.05 \to 5.6) \times 10^{-3}$~\\

\hline
\end{tabular}
\end{center}
\end{table}

\subsection{$B_s \to \mu^\mp e^\pm$ process}
The constraints on LQ couplings obtained from the branching ratio of the lepton flavor violating (LFV) $B_s \to \mu^\mp e^\pm$ process is discussed in this subsection. In the SM,  the LFV decay modes occur at loop level with the presence of tiny neutrinos in one of the loops or proceed via box diagrams. However, these processes can occur at tree level in the vector LQ model. The present experimental upper bound on the branching ratio of $B_s \to \mu^\mp e^\pm$ process is \cite{pdg}
\bea \label{bsmue-EXP}
{\rm BR}(B_s \to \mu^\mp e^\pm) < 1.1 \times 10^{-8}.
\eea
In the presence of $V^1(3,1,2/3)$ vector LQ, the branching ratio of $B_s \to \mu^- e^+$ decay mode is 
\bea
 {\rm BR}(B_s\to \mu^- e^+)&=& \tau_{B_s} \frac{G_F^2\alpha^2 M_{B_s}^5f_{B_s}^2 |V_{tb}V_{ts}^*|^2}{64\pi^3}  \left(1-\frac{m_\mu^2}{M_{B_s}^2}\right)^2   \Bigg[\left| \frac{ m_\mu }{M_{B_s}^2} \left(G_9^{\rm LQ}-G_9^{\prime \rm LQ}\right) + \frac{G_S^{\rm LQ}-G_S^{\prime \rm LQ}}{m_b}  \right|^2 \nn \\ && +\left|\frac{m_\mu}{M_{B_s}^2}\left(G_{10}^{\rm LQ}-G_{10}^{\prime \rm LQ}\right) + \frac{G_P^{\rm LQ}-G_P^{ \prime \rm LQ}}{m_b}\right|^2\Bigg] \, ,    \label{bsmue1} 
\eea
and the branching ratio of $B_s \to \mu^+ e^-$ decay process is given by 
\bea
 {\rm BR}(B_s\to  e^- \mu^+)&=& \tau_{B_s} \frac{G_F^2\alpha^2 M_{B_s}^5f_{B_s}^2 |V_{tb}V_{ts}^*|^2}{64\pi^3}  \left(1-\frac{m_\mu^2}{M_{B_s}^2}\right)^2   \Bigg[\left| -\frac{ m_\mu }{M_{B_s}^2} \left(H_9^{\rm LQ}-H_9^{\prime \rm LQ}\right) + \frac{H_S^{\rm LQ}-H_S^{\prime \rm LQ}}{m_b}  \right|^2 \nn \\ && +\left|\frac{m_\mu}{M_{B_s}^2}\left(H_{10}^{\rm LQ}-H_{10}^{\prime \rm LQ}\right) + \frac{H_P^{\rm LQ}-H_P^{ \prime \rm LQ}}{m_b}\right|^2\Bigg] \,, \label{bsmue2} 
\eea
where the mass of electron is neglected.
 Here the new $G(H)_a^{(\prime)\rm LQ}~(a=9,10,S, P)$  coefficients  have similar expression as Eqns. (\ref{u1c10np},\ref{u1c10pnp}, \ref{u1csnp},\ref{u1cspnp}) with the replacement of LQ couplings $(g_i)_{sl}(g_j)_{bl}^* \to (g_i)_{se}(g_j)_{b\mu}^*$, where $(i,j=L,R)$ for $G_a^{(\prime)\rm LQ}$  and  $(g_i)_{sl}(g_j)_{bl}^* \to (g_i)_{s\mu}(g_j)_{be}^*$ for $H_a^{(\prime)\rm LQ}$ coefficients. 

The  total branching ratio of $B_s \to \mu^\mp e^\pm$ process is
\bea 
{\rm BR}(B_s \to \mu^\mp e^\pm) = {\rm BR}(B_s \to \mu^- e^+)+{\rm BR}(B_s \to \mu^+ e^-).
\eea
For chiral LQ, only $G(H)_{9,10}^{(\prime)\rm LQ}$ coefficients will be present. Now using the experimental upper limit of the branching ratio  (\ref{bsmue-EXP}), we obtain the  constraint on LQ couplings as
\bea
|(g_L)_{se}(g_L)_{b\mu}^*| < 2.83 \times 10^{-2}.
\eea
Now neglecting the $V\pm A$ couplings, the constraint on $(G_S^{\rm LQ} \pm G_S^{\prime \rm LQ})$ coefficient is shown in Fig. 1. From the figure, we find the allowed range
for the above combinations of Wilson coefficients as
\bea
|G_S^{\rm LQ}-G_S^{\prime \rm LQ}| \leq 0.3,~~~ |G_S^{\rm LQ}+G_S^{\prime \rm LQ}| \leq 0.3\,. 
\eea
\begin{figure}[h]
\centering
\includegraphics[scale=0.55]{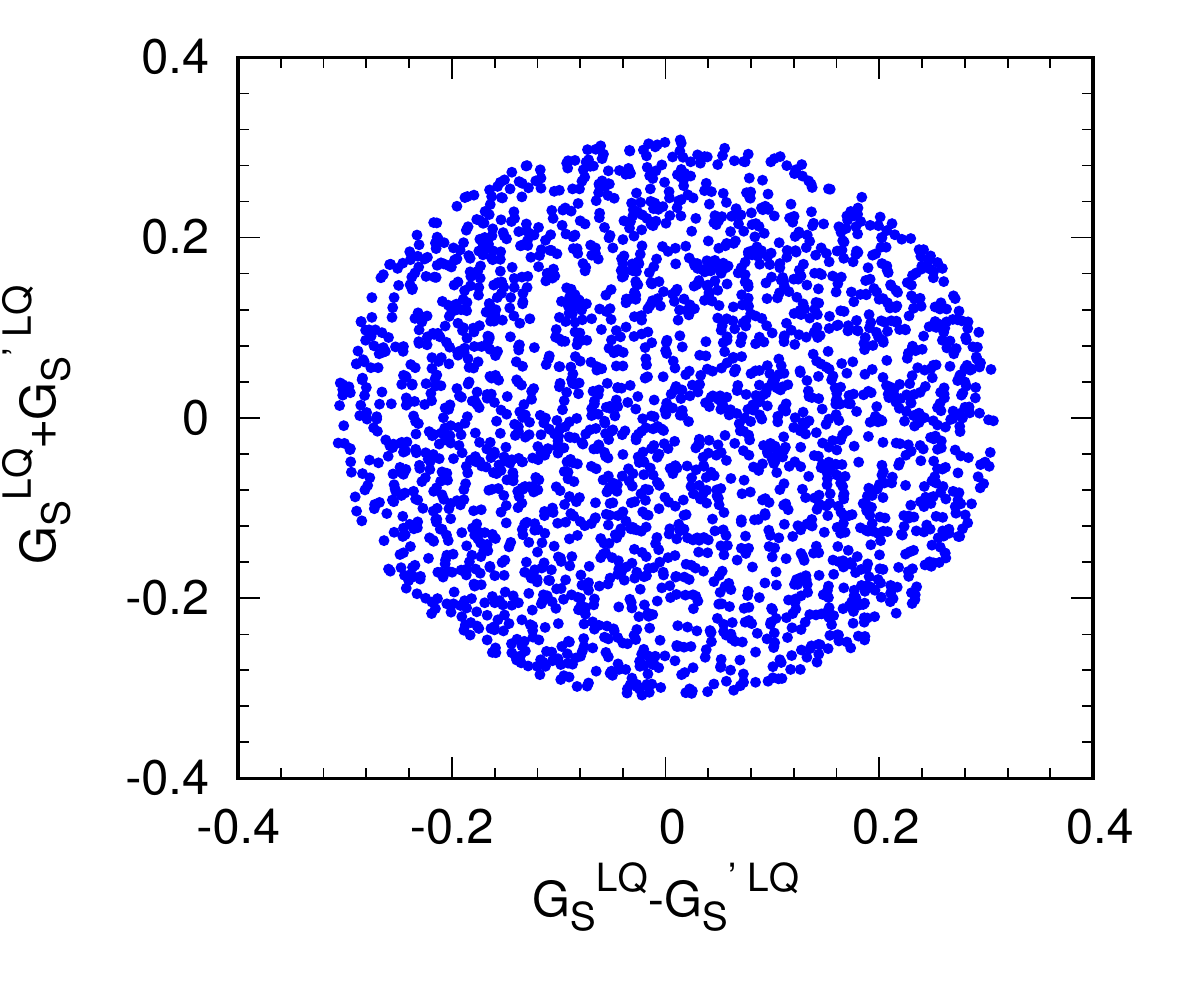}
\caption{The constraint on $G_S^{\rm LQ} \pm G_S^{\prime \rm LQ}$ couplings obtained from the branching ratio of $B_s \to \mu^- e^+$ process.}
\end{figure}
\section{$\bar B \to \bar K^* l^+ l^-$ processes}
In this section, we present the theoretical framework to calculate the branching ratios  for the rare semileptonic  $\bar B \to \bar K^* l^+ l^-$ processes.   Furthermore, the dileptons present in these processes 
allow one to formulate several useful observables which
can be  used  to probe and discriminate different
scenarios of NP.
 The full four body angular distribution of the  
$\bar{B} \rightarrow \bar{K}^{* 0}\left(\rightarrow K^-\pi^+\right) l^+ l^-$ decay processes  can be described  by four independent kinematic variables, 
 $q^2$  and the three angles $\theta_{K^*}, \theta_l$ and $  \phi$. Here we assume that, $\bar{K}^{* 0}\rightarrow K^-\pi^+$ is  on the mass shell. The differential decay distribution of these processes  with respect to the four independent   variables are given  as \cite{kstar-br-1, kstar-br-2, kstar-br-3}
 \begin{equation}
 \frac{d^4\Gamma}{dq^2~ d\cos\theta_l ~d\cos\theta_{K^*}~ d\phi} = \frac{9}{32\pi} J\left(q^2, \theta_l, \theta_{K^*}, \phi\right),
 \end{equation}
 where the lepton spins have been summed over. 
 Here $q^2$ is the lepton-pair invariant mass square, $\theta_l$ is  the angle between the negatively charged lepton and 
the $\bar{B}$ in the $l^+ l^-$ frame, $\theta_{K^*}$ is defined as  the angle between $K^-$ and $\bar{B}$ in the $K^-\pi^+$ center 
of mass frame and $\phi$ is  the angle between the normals of the $K^-\pi^+$ and the dilepton  planes.
The physically allowed  regions  of these variables in the phase space are  given by 
 \begin{equation}
 4m^2_l \leqslant q^2 \leqslant \left(m_B - m_{K^*}\right)^2,\hspace{0.6cm} -1\leqslant \cos\theta_l \leqslant 1,\hspace{0.6 cm} -1
\leqslant \cos\theta_{K^*} \leqslant 1,\hspace{0.7cm} 0\leqslant \phi \leqslant 2\pi,
 \end{equation}
 where $m_B ~(m_{K^*}$) and  $m_l$ are respectively the masses of $B~(K^*)$ meson and charged-lepton. 
The explicit  dependence of the decay distribution on the above  three angles,  (i.e., the  dependence $J\left(q^2, \theta_l, \theta_{K^*}, \phi\right)$ function) can be written as
\beqa
 J\left(q^2, \theta_l, \theta_{K^*}, \phi\right) &= & J^s_1 \sin^2\theta_{K^*} + J^c_1 \cos^2\theta_{K^*} + \left(J^s_2 \sin^2\theta_{K^*} 
+ J^c_2 \cos^2\theta_{K^*}\right) \cos2\theta_l \nn\\
& +& J_3 \sin^2\theta_{K^*} \sin^2\theta_l \cos2\phi + J_4 \sin2\theta_{K^*} \sin2\theta_l \cos\phi 
  +  J_5 \sin2\theta_{K^*} \sin\theta_l \cos\phi\nn\\
& +&(J_6^s \sin^2\theta_{K^*} +J_6^c \cos^2\theta_{K^*})\cos\theta_l
 + J_7 \sin2\theta_{K^*} \sin\theta_l \sin\phi 
\nn\\ 
& +&  J_8 \sin2\theta_{K^*} \sin2\theta_l \sin\phi + J_9 \sin^2\theta_{K^*} \sin^2\theta_l \sin2\phi\;,
\eeqa
 where the coefficients $J_i^{(a)} = J_i^{(a)}\left(q^2\right)$ for $i = 1,....,9$ and $a = s,c$ are functions of the dilepton invariant mass. The complete expression for these coefficients  in terms of the transversity amplitudes $A_0$, $A_\parallel$, $A_\perp$, and $A_t$ can be found in the Ref. \cite{kstar-br-1, kstar-br-2, kstar-fl-2}. 
 
After performing the  integration over all the  angles,  the decay rate of  $\bar{B} \rightarrow \bar{K}^* l^+ l^-$ processes with respect to $q^2$  is given by  \cite{kstar-br-1}
 \begin{equation}
 \frac{d\Gamma}{dq^2} = \frac{3}{4} \left(J_1 - \frac{J_2}{3}\right),
 \end{equation}
where $J_i = 2J_i^s + J_i^c$. 

Previously LHCb had measured the LNU parameter   in the low $q^2$, i.e., $(1 \leq q^2 \leq 6) ~ {\rm GeV}^2$ region  of $B \to K l^+ l^-$ process  as \cite{RK-exp} 
\bea
R_K^{\rm LHCb} = \frac{{\rm BR}(B^+ \to K^+ \mu^+ \mu^-)}{{\rm BR}(B^+ \to K^+ e^+ e^-)}=  0.745^{+0.090}_{-0.074} \pm 0.036
\eea
which has  $2.6 \sigma$ deviation 
from the corresponding SM result $R_K^{\rm SM} = 1.0003 \pm 0.0001$ \cite{RK-SM}. Recently, LHCb collaboration has measured  analogous   lepton flavour universality  violating  parameter, $R_{K^*}$, in the $\bar B \to \bar  K^* l^+ l^-$ processes in two different bins,  which also have around $2 \sigma$ deviations from the corresponding SM values as presented in Table-I.
Besides the branching ratios and the  $R_{K^*}$ parameter, there are many observables associated with $\bar B \to \bar K^* l^+ l^-$ processes which  could be sensitive to new physics. The interesting observables  are 
\begin{enumerate}

\item The zero crossing of the  forward-backward asymmetry, which is defined as \cite{kstar-br-1} 
\beqa
  A_{FB}\left(q^2\right) & = & \left[ \int_{-1}^0 d\cos\theta_l \frac{d^2\Gamma}{dq^2 d\cos\theta_l}
 - \int_{0}^1 d\cos\theta_l \frac{d^2\Gamma}{dq^2 d\cos\theta_l}\right] \Big{/}
  \frac{d\Gamma}{d q^2}\nn \\&  =& -\frac{3}{8} \frac{J_6}{d\Gamma/dq^2}\;.
   \eeqa
   
   \item 
   The longitudinal and transverse polarization fractions of the $K^*$ meson, in terms of the angular coefficients $(J_i)$  can be written as \cite{kstar-fl-1, kstar-fl-2}
\begin{equation}
 F_L\left(q^2\right) = \frac{3J_1^c-J_2^c}{4d\Gamma/dq^2}\;, \hspace{1.5 cm}
 F_T\left(q^2\right) = 1-F_L(q^2)\,.
 \end{equation}
 
 \item 
 The form factor independent (FFI) optimized observables $P_i$'s, where $i=1, ..,6,8$ are given as \cite{kstar-p1}
  \beqa
 P_1\left(q^2\right) &=& \frac{J_3}{2J_2^s}\;, \hspace{1.5cm}  P_2\left(q^2\right)  =  \beta_l\frac{J^s_6}{8J_2^s}\;,\hspace{1.5cm}
 P_3\left(q^2\right)  = - \frac{J_9}{4J_2^s}\;,\nn\\
 P_4\left(q^2\right) & = & \frac{\sqrt{2}J_4}{\sqrt{-J_2^c\left(2J_2^s-J_3\right)}}\;,\hspace{1.2cm}
 P_5\left(q^2\right) =  \frac{\beta_l J_5}{\sqrt{-2J_2^c\left(2J_2^s+J_3\right)}}\;,\nn\\
 P_6\left(q^2\right) & = & - \frac{\beta_l J_7}{\sqrt{-2J_2^c\left(2J_2^s-J_3\right)}}\;, \hspace{1cm}P_8\left(q^2\right)  =  - \frac{\beta_l J_8}{\sqrt{-2J_2^c\left(2J_2^s-J_3\right)}}\;.
 \eeqa
 
 \item 
 In order to interpret the  LHCb measurements more precisely, a  slightly modified set of clean observables $P_{4,5,6,8}^\prime$, which related to $P_{4,5,6, 8}$ are defined as \cite{kstar-p4p}
\bea
&&P_{4}^\prime  \equiv P_{4} \sqrt{1-P_1} = \frac{J_{4}}{\sqrt{-J_2^c J_2^s}}\;,\nn \\ 
&&P_{5}^\prime  \equiv P_{5} \sqrt{1+P_1} = \frac{J_{5}}{2\sqrt{-J_2^c J_2^s}}\;,\nn \\ 
&&P_{6, 8}^\prime \equiv P_{6, 8} \sqrt{1-P_1} = -\frac{J_{7, 8}}{2\sqrt{-J_2^c J_2^s}}\;.
\eea

 \item 
To confirm the existence of the violation of lepton universality,  one can define additional LFUV  observables as \cite{kstar-Q4}
\bea
&&Q_{F_L} =F_L^\mu - F_L^e , ~~~~~Q_{F_T} =F_T^\mu - F_T^e , \\
&&Q_i= P_i^{\mu } - P_i^{e}, ~~~~~~~~B_i = \frac{J_i^\mu}{J_i^e} -1.
\eea
where $P_i$'s should be replaced by $P_i^\prime$ for $Q_{4,5,6,8}$.
\end{enumerate}

\begin{figure}[h]
\centering
\includegraphics[scale=0.45]{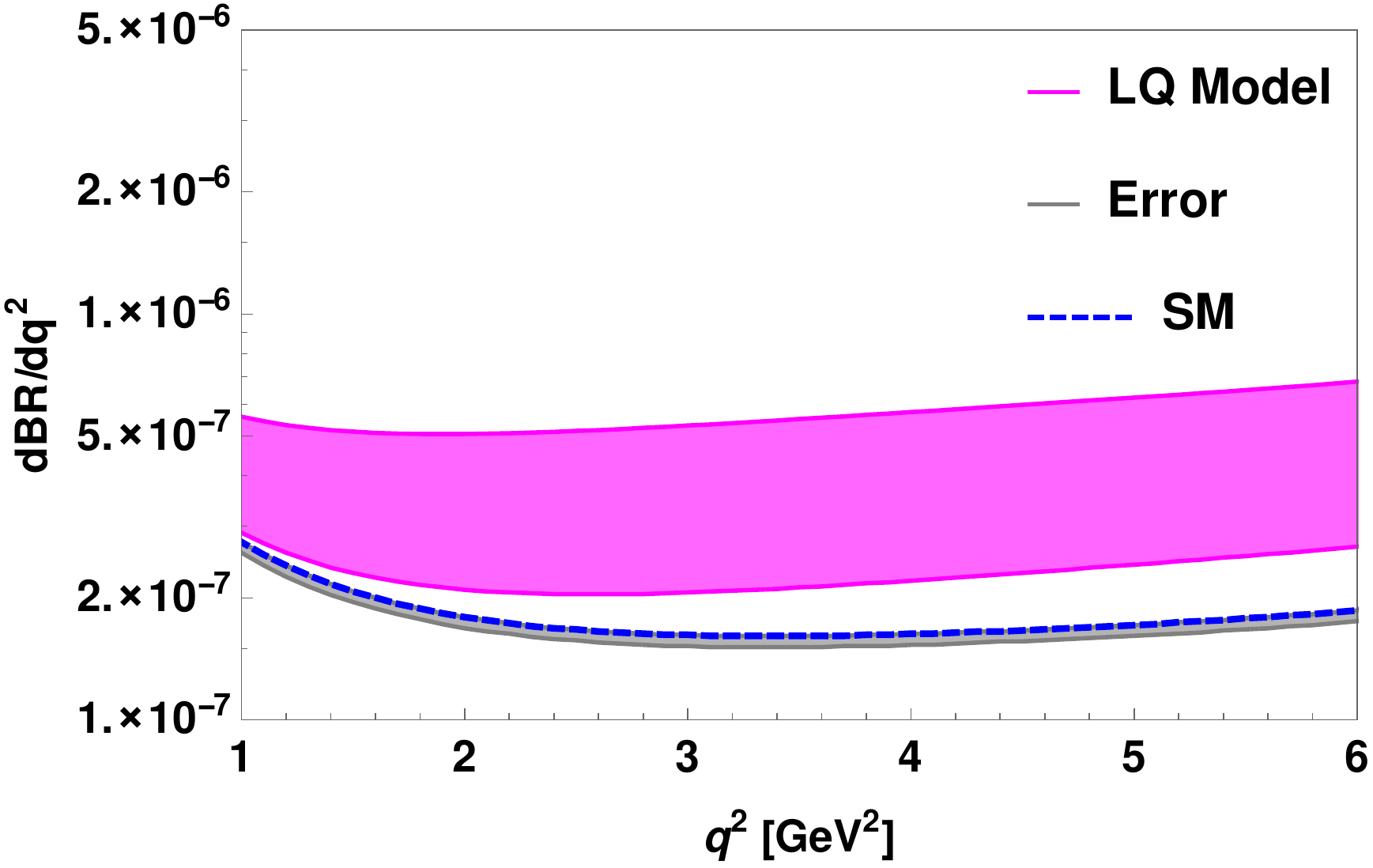}
\quad
\includegraphics[scale=0.45]{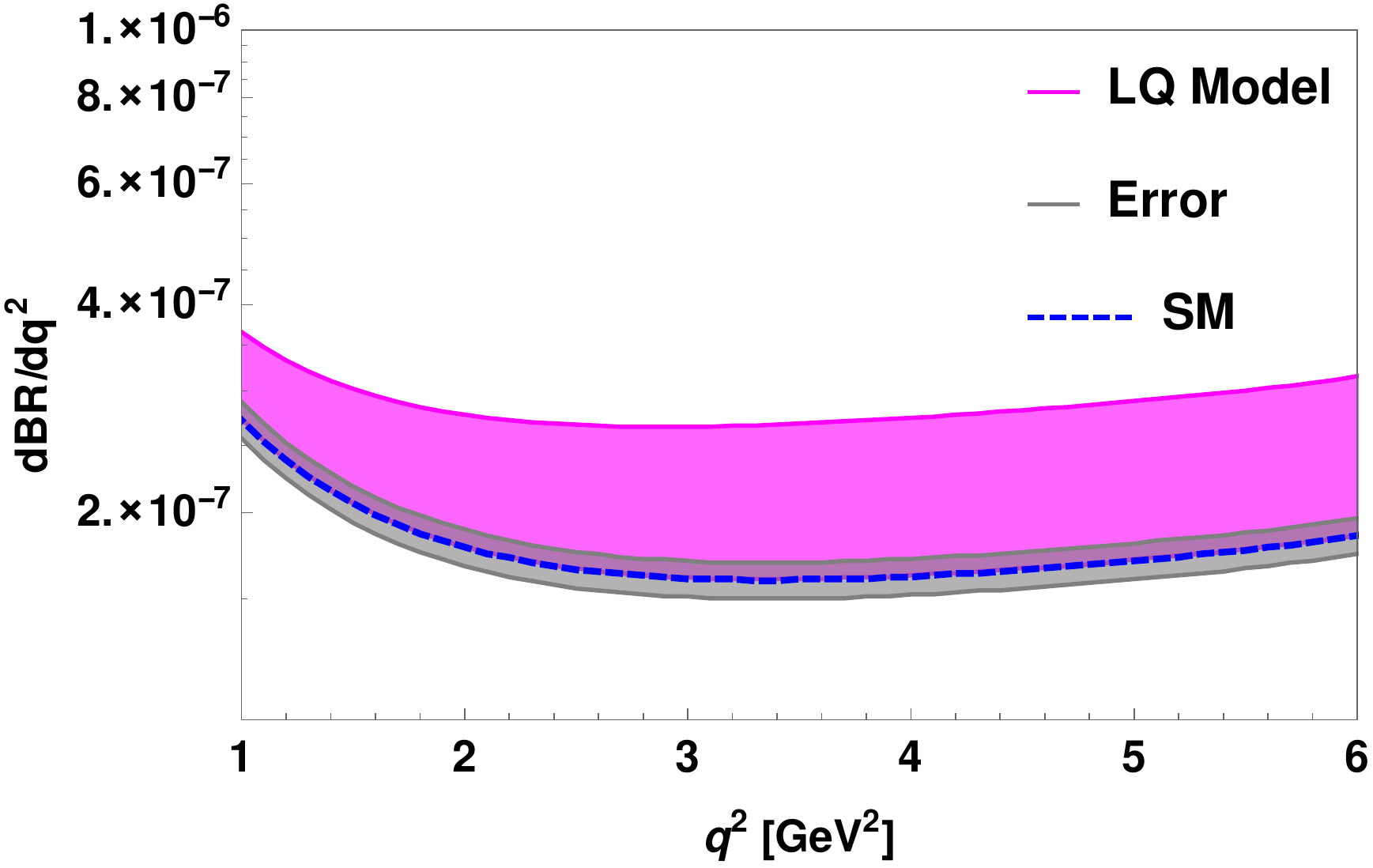}
\caption{The differential branching ratios of $\bar B \to \bar K^* e^+ e^-$ (left panel) and $\bar B \to \bar K^* \mu^+ \mu^-$ (right panel) processes with respect to the $q^2$  in the vector LQ model. Here the magenta bands represent the LQ contributions and the  dotted lines are for the SM. The theoretical uncertainties arising due to the SM  input parameters are shown as grey bands.}
\end{figure}

\begin{figure}[h]
\centering
\includegraphics[scale=0.45]{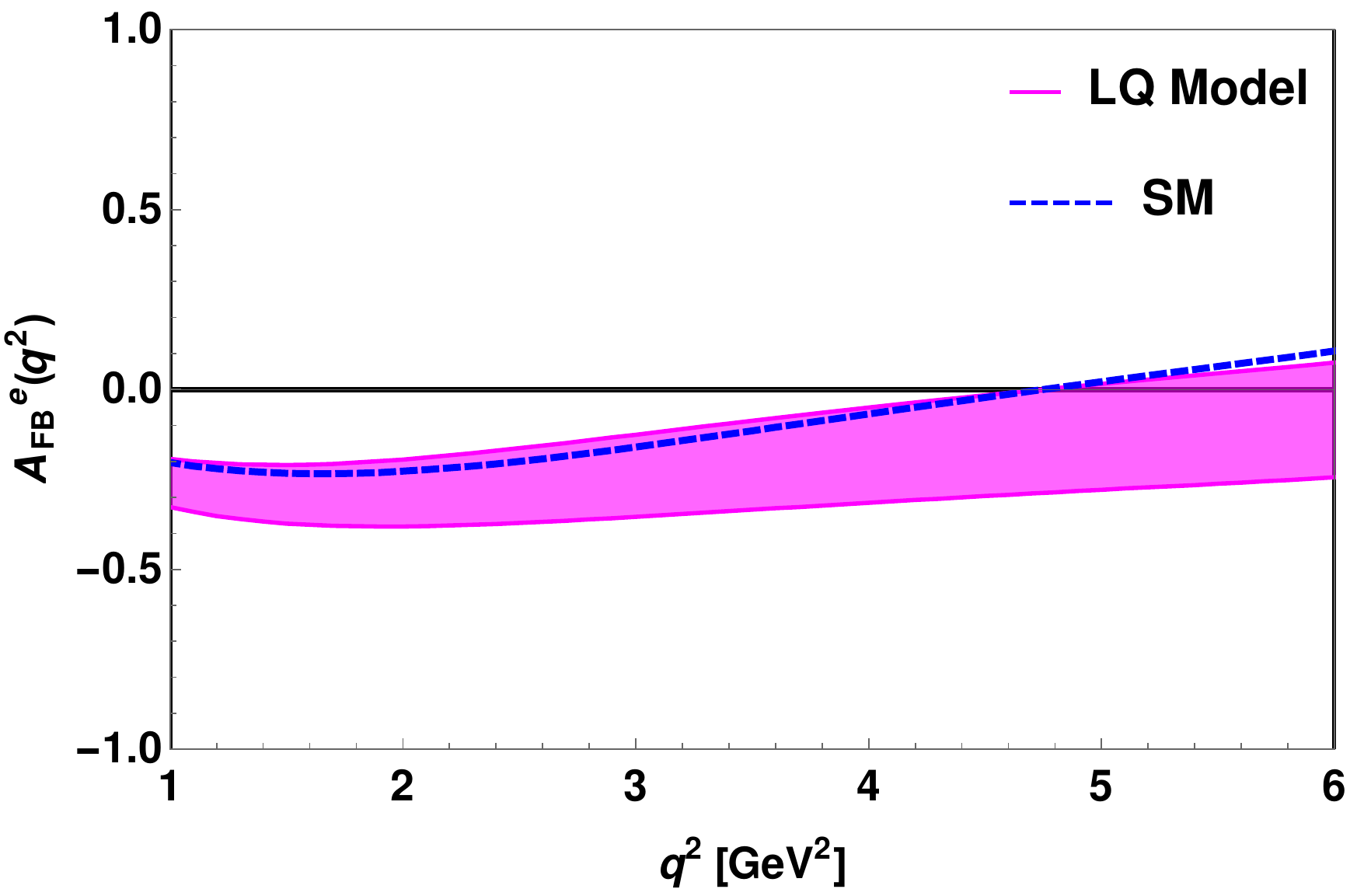}
\quad
\includegraphics[scale=0.45]{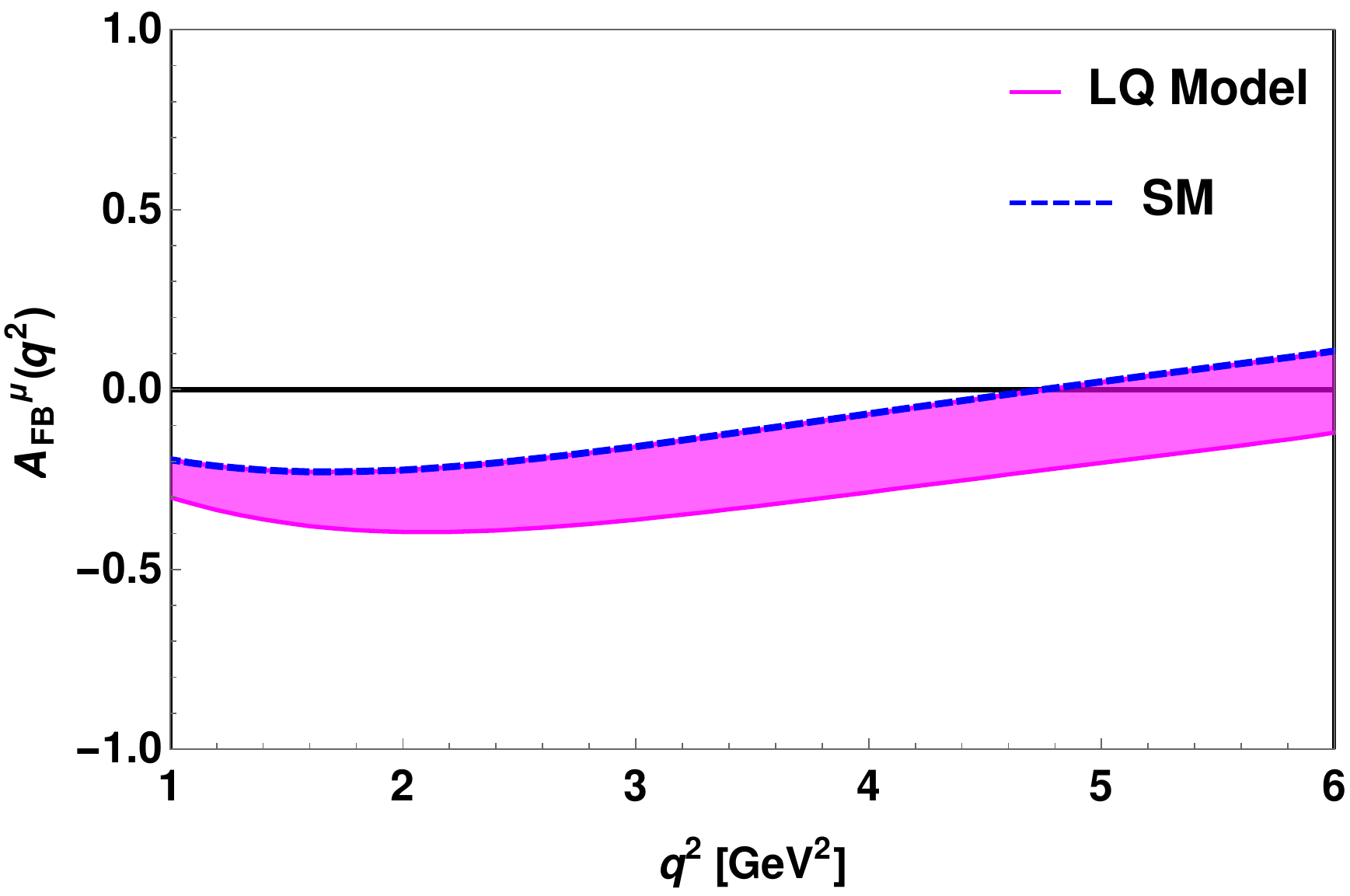}
\caption{The $q^2$ variations of the forward-backward asymmetries  of  $\bar B \to \bar K^* e^+ e^-$ (left panel) and $\bar B \to \bar K^* \mu^+ \mu^-$ (right panel) processes  in the vector LQ model. }
\end{figure}

\begin{figure}[h]
\centering
\includegraphics[scale=0.45]{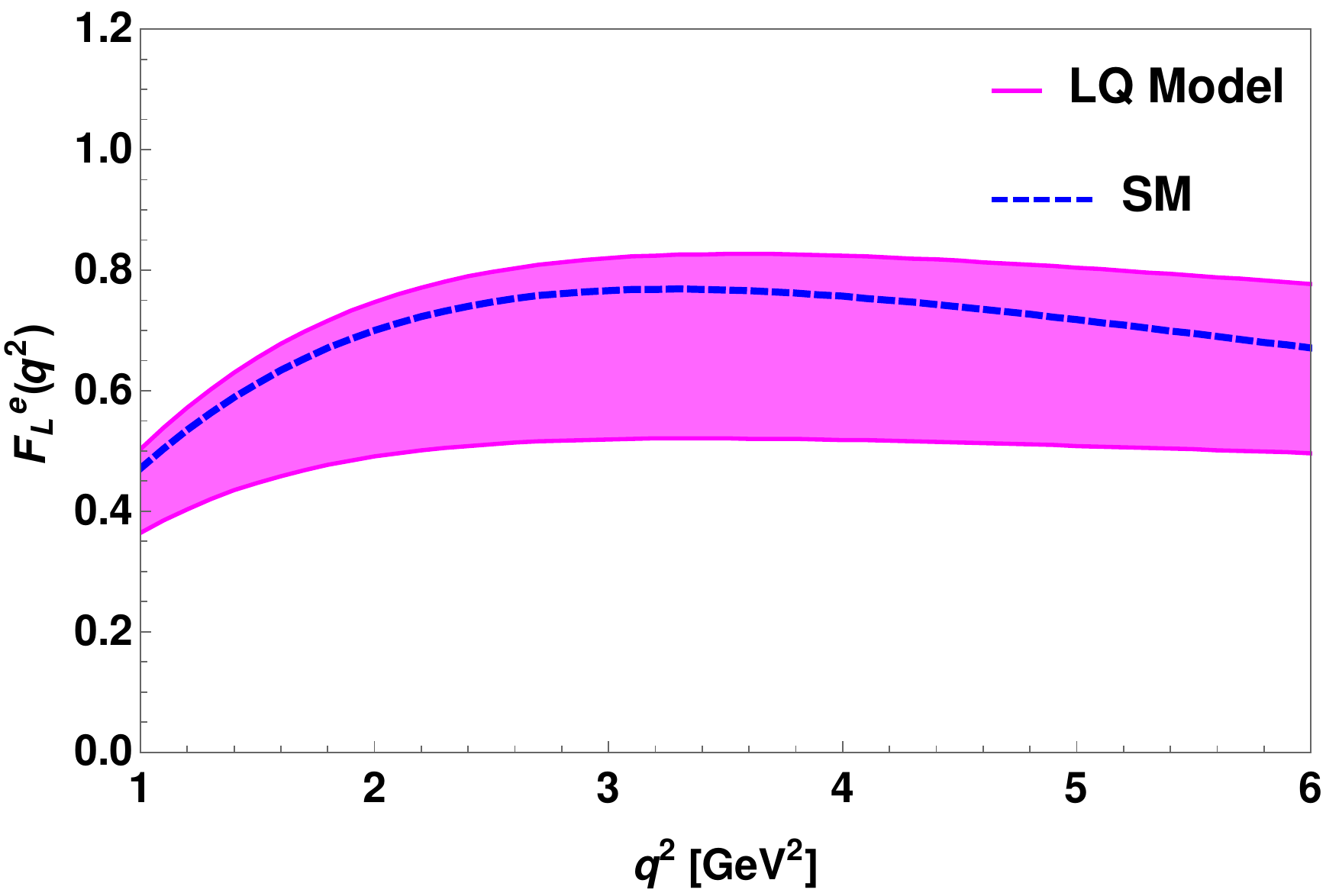}
\quad
\includegraphics[scale=0.45]{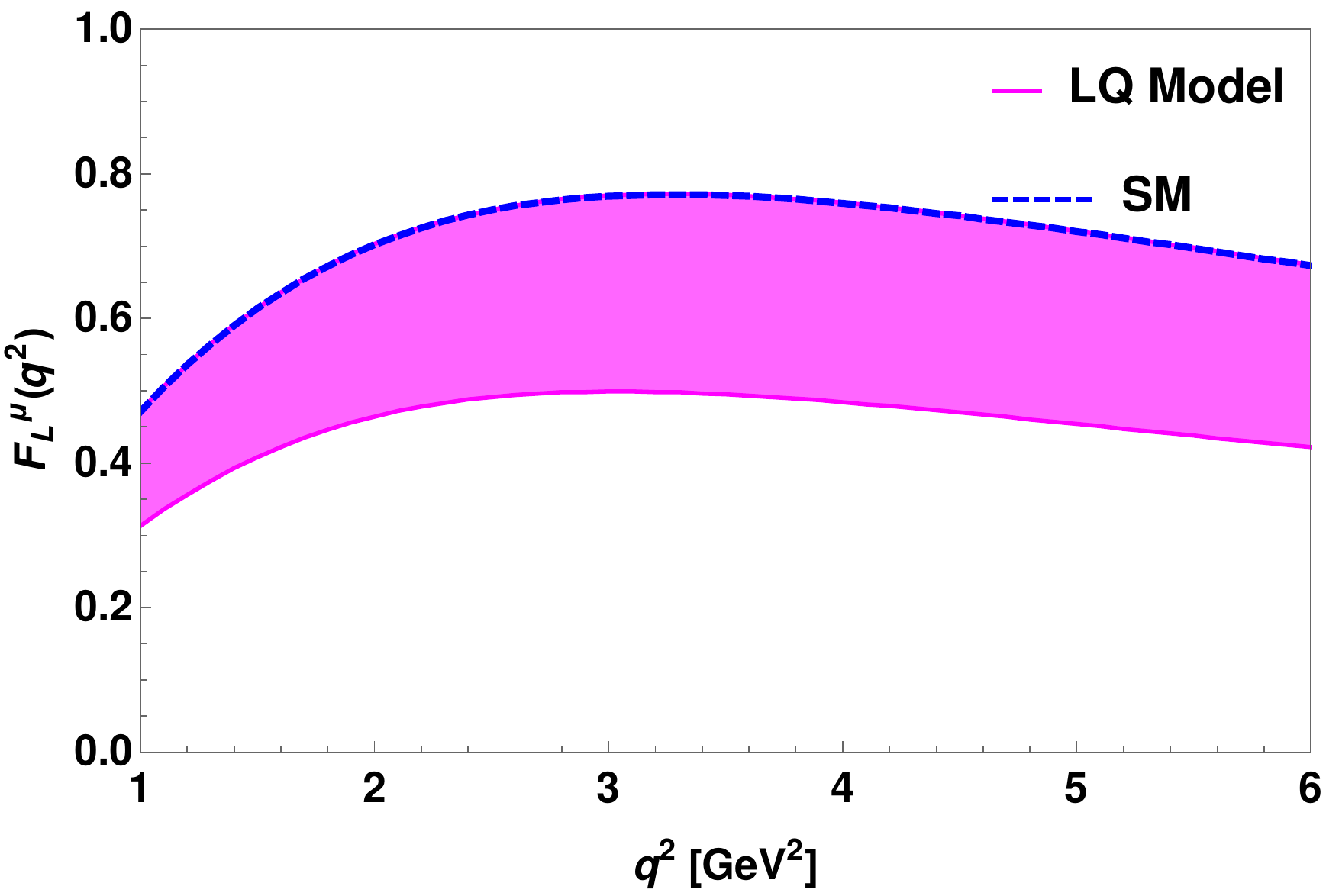}
\caption{The $q^2$ variations of the longitudinal polarizations  of  $\bar B \to \bar K^* e^+ e^-$ (left panel) and $\bar B \to \bar K^* \mu^+ \mu^-$ (right panel) processes  in the vector LQ model. }
\end{figure}

\begin{figure}[h]
\centering
\includegraphics[scale=0.45]{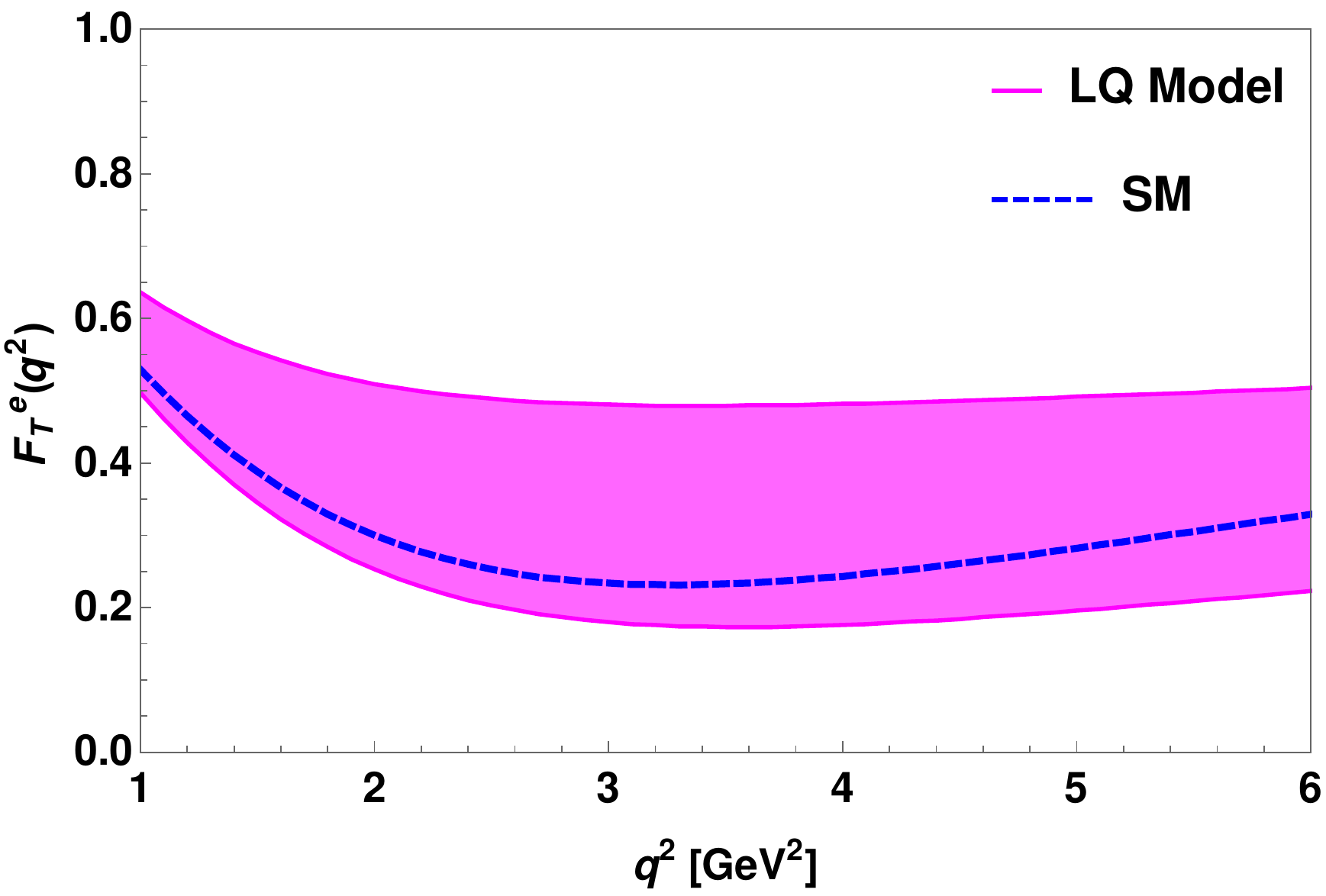}
\quad
\includegraphics[scale=0.45]{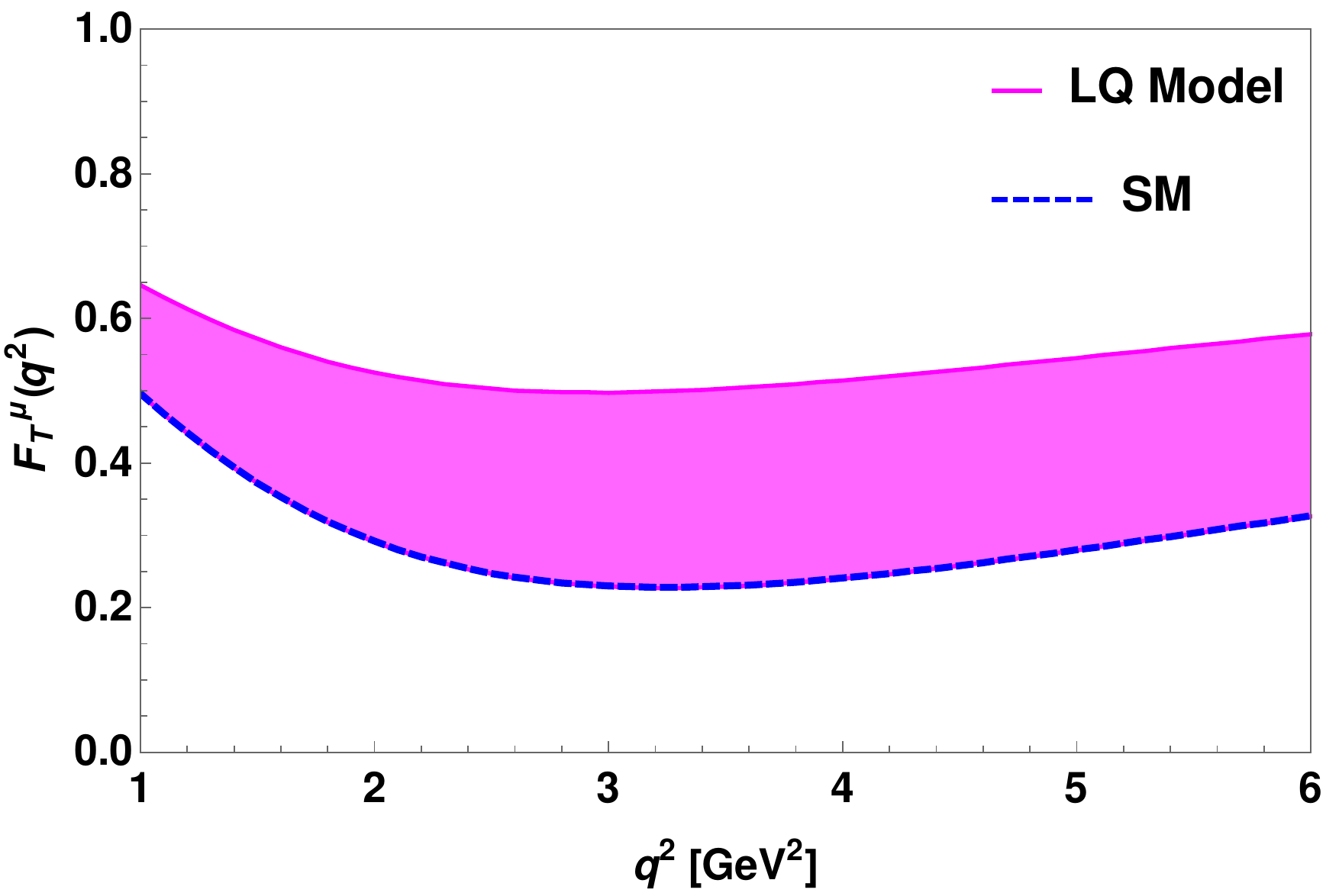}
\caption{The $q^2$ variations of the transverse polarizations of  $\bar B \to \bar K^* e^+ e^-$ (left panel) and $\bar B \to \bar  K^* \mu^+ \mu^-$ (right panel) processes  in the vector LQ model. }
\end{figure}

\begin{figure}[h]
\centering
\includegraphics[scale=0.5]{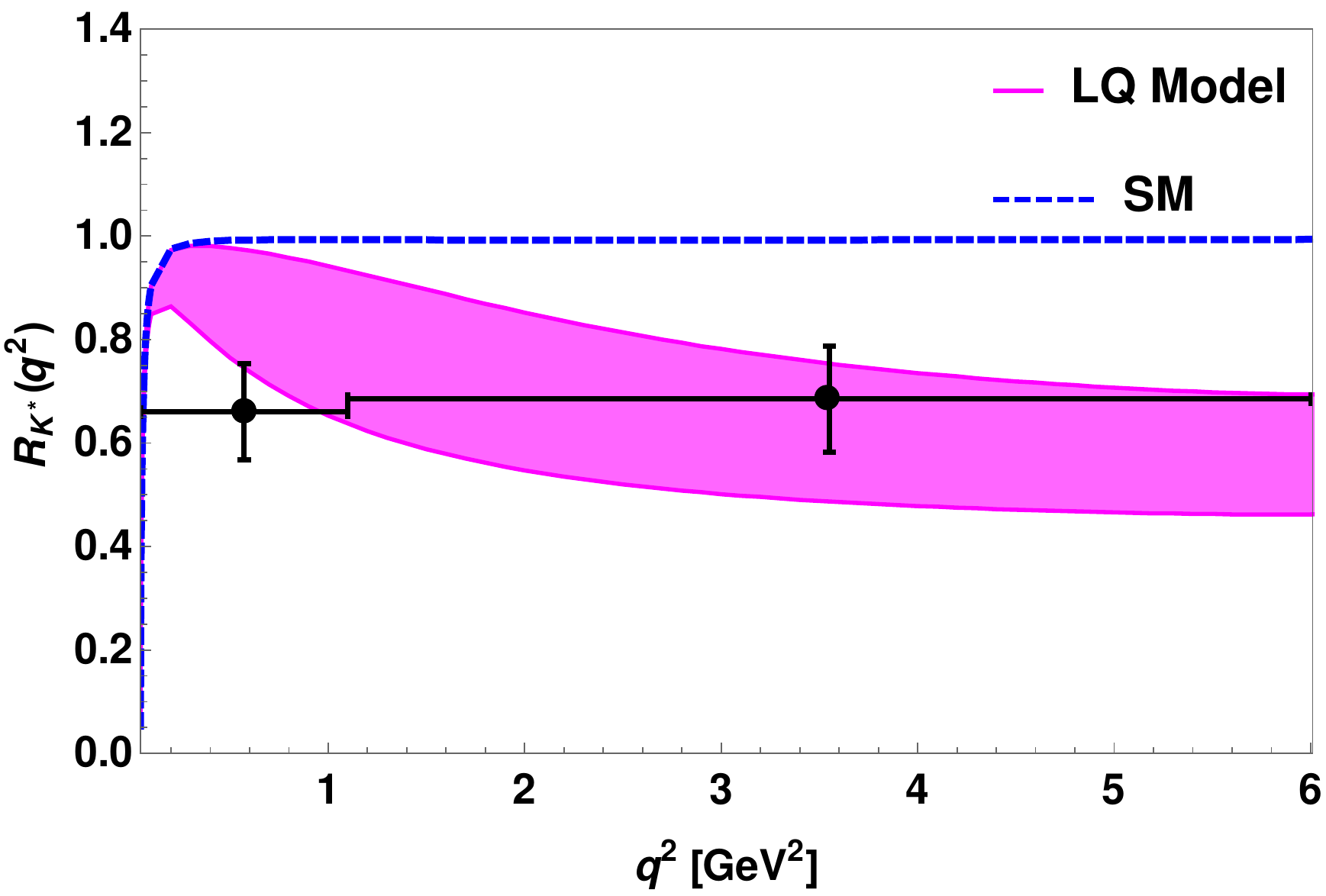}
\caption{The  variations of $R_{K^*}(q^2)$  in the $q^2\in [0.045,6.0]~{\rm GeV}^2$  regions  in the vector LQ model. }
\end{figure}

\begin{table}[htb]
\begin{center}
\caption{The predicted  integrated values of the branching ratio, forward-backward asymmetry and lepton polarization asymmetry   with respect to  low $q^2$  for the $\bar B \rightarrow \bar  K^* l^+ l^-$ processes in the SM and the vector LQ model. }
\begin{tabular}{|c | c | c| }
\hline
 Observables & SM prediction & Values in  LQ model \\
 \hline
 \hline
{\rm BR}($\bar B \rightarrow \bar K^*  e^+ e^-$) & $(8.97 \pm 0.49~({\rm CKM}) \pm 0.23~({\rm form~factor})) \times 10^{-7}$ & $(1.155 \to 2.882) \times 10^{-6}$  \\
 
 $\langle A_{FB}^e \rangle$ & $-0.084 \pm 0.005$  & $-(0.314 \to 0.064)$   \\
 
 $\langle F_L^e \rangle$ & $0.703 \pm 0.042$ & $0.5 \to 0.76$   \\
 
 $\langle F_T^e \rangle$ & $0.297 \pm 0.018$ & $0.24 \to 0.5$ \\

 \hline
 
{\rm BR}($\bar B \rightarrow \bar  K^*   \mu^+ \mu^-$) & $(8.9 \pm 0.48~({\rm CKM}) \pm ~0.22~({\rm form~factor})) \times 10^{-7}$ & $(0.892 \to 1.45) \times 10^{-6}$  \\
 
$\langle A_{FB}^\mu\rangle$ &$-0.082 \pm 0.0049$ & $-(0.28 \to 0.083)$ \\

$\langle F_L^\mu \rangle$ & $0.71 \pm 0.043$ & $0.46 \to 0.71$  \\
$\langle F_T^\mu \rangle$ & $0.29 \pm 0.017$ & $0.29 \to 0.54$  \\

 \hline
\end{tabular}
\end{center}
\end{table}



\begin{table}[htb]
\begin{center}
\caption{The predicted  integrated values of the lepton non-universality $(R_{K^*})$ parameter in  the LQ model. }
\begin{tabular}{|c | c | c|  }
\hline
 Observables & SM prediction & Values in  LQ model  \\
 \hline
 \hline

 $\langle R_{K^*} \rangle |_{q^2 \in [0.045,1.1]}$ & $0.913 $ & $0.65 \to 0.9$ \\
 
 $\langle R_{K^*} \rangle|_{q^2 \in [1.1, 6.0]}$ & $0.9926 $  & $0.5 \to 0.73$ \\

 \hline
\end{tabular}
\end{center}
\end{table}


\begin{figure}[h]
\centering

\includegraphics[scale=0.45]{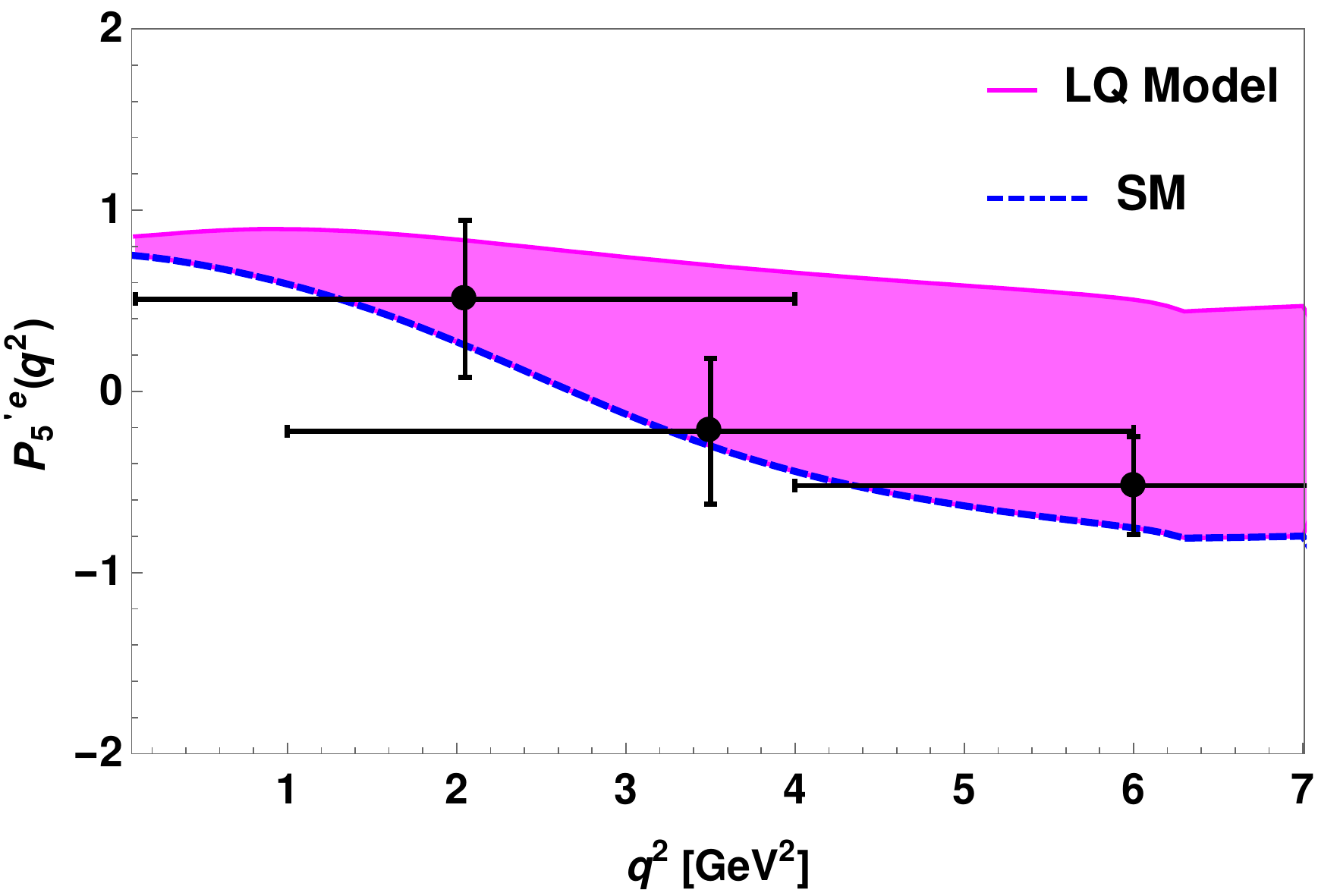}
\quad
\includegraphics[scale=0.45]{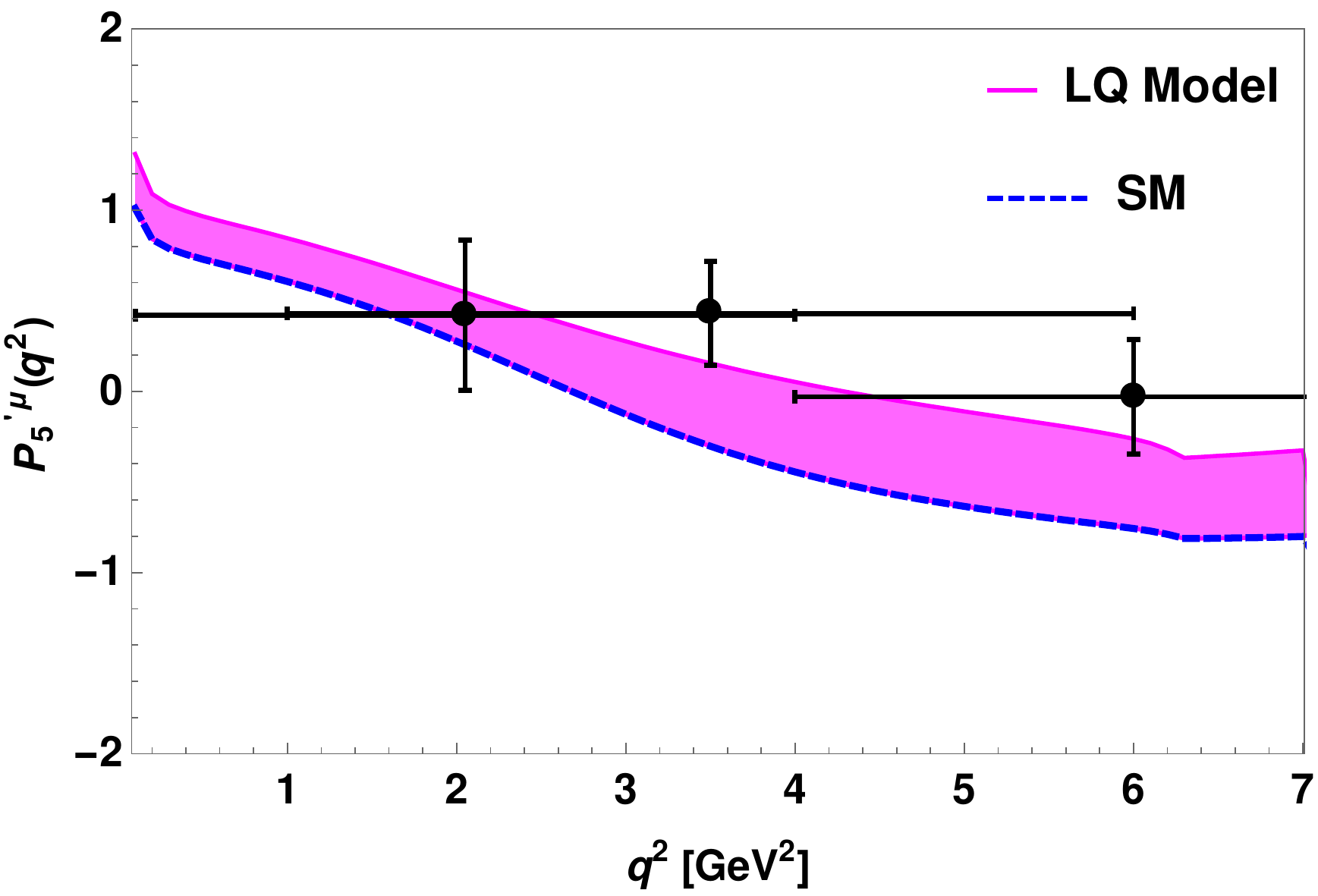}\\
\caption{The plot in the left panel represent the $P_5^\prime (q^2)$ observable for $\bar B \rightarrow \bar  K^* e^+ e^-$ precess in the vector LQ model. The corresponding   plot for $\bar B \rightarrow \bar K^* \mu^+ \mu^-$ process is shown in the right panel.}
\end{figure}

\begin{figure}[htb]
\centering
\includegraphics[scale=0.45]{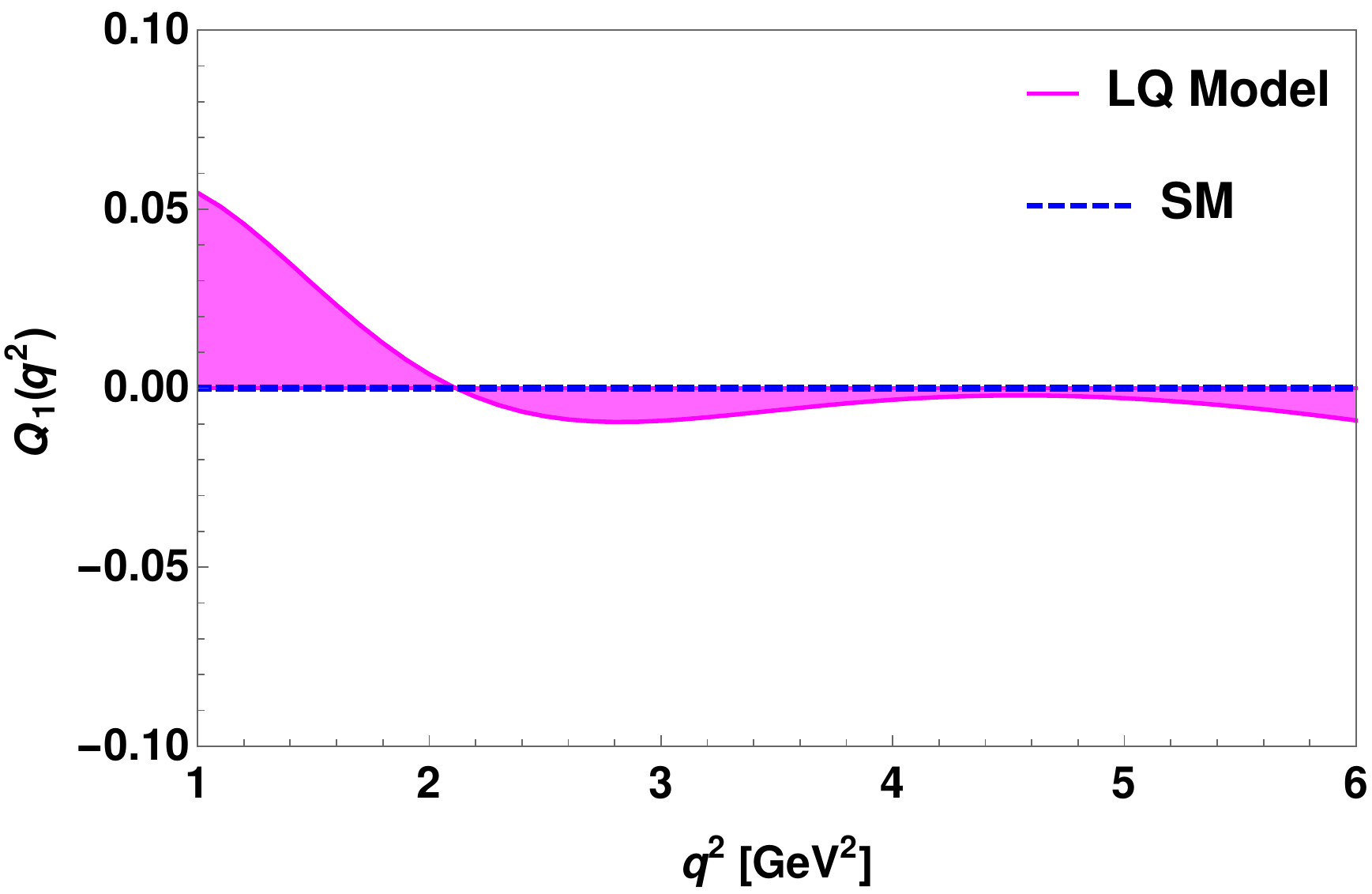}
\quad
\includegraphics[scale=0.45]{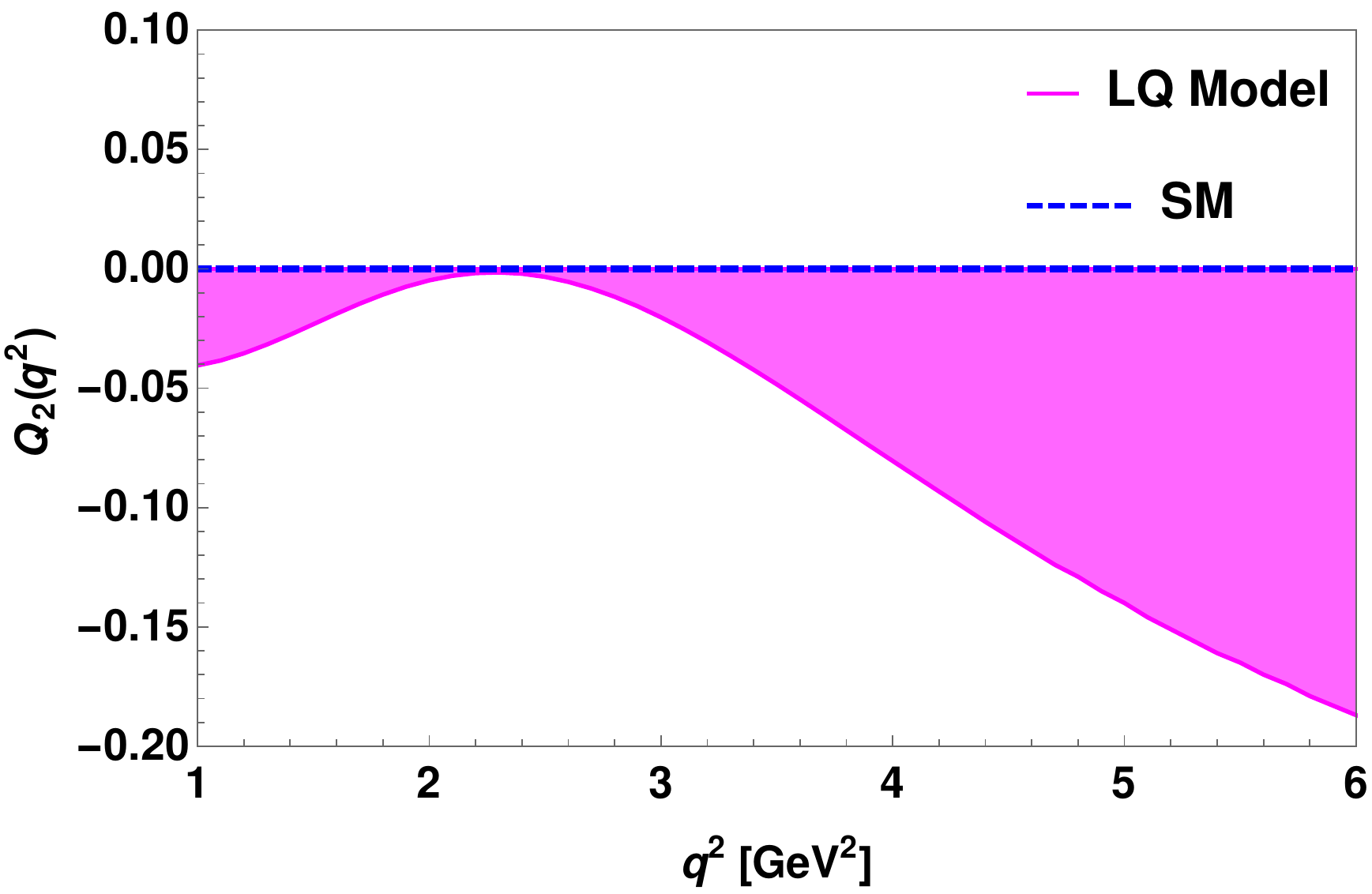}\\
\includegraphics[scale=0.45]{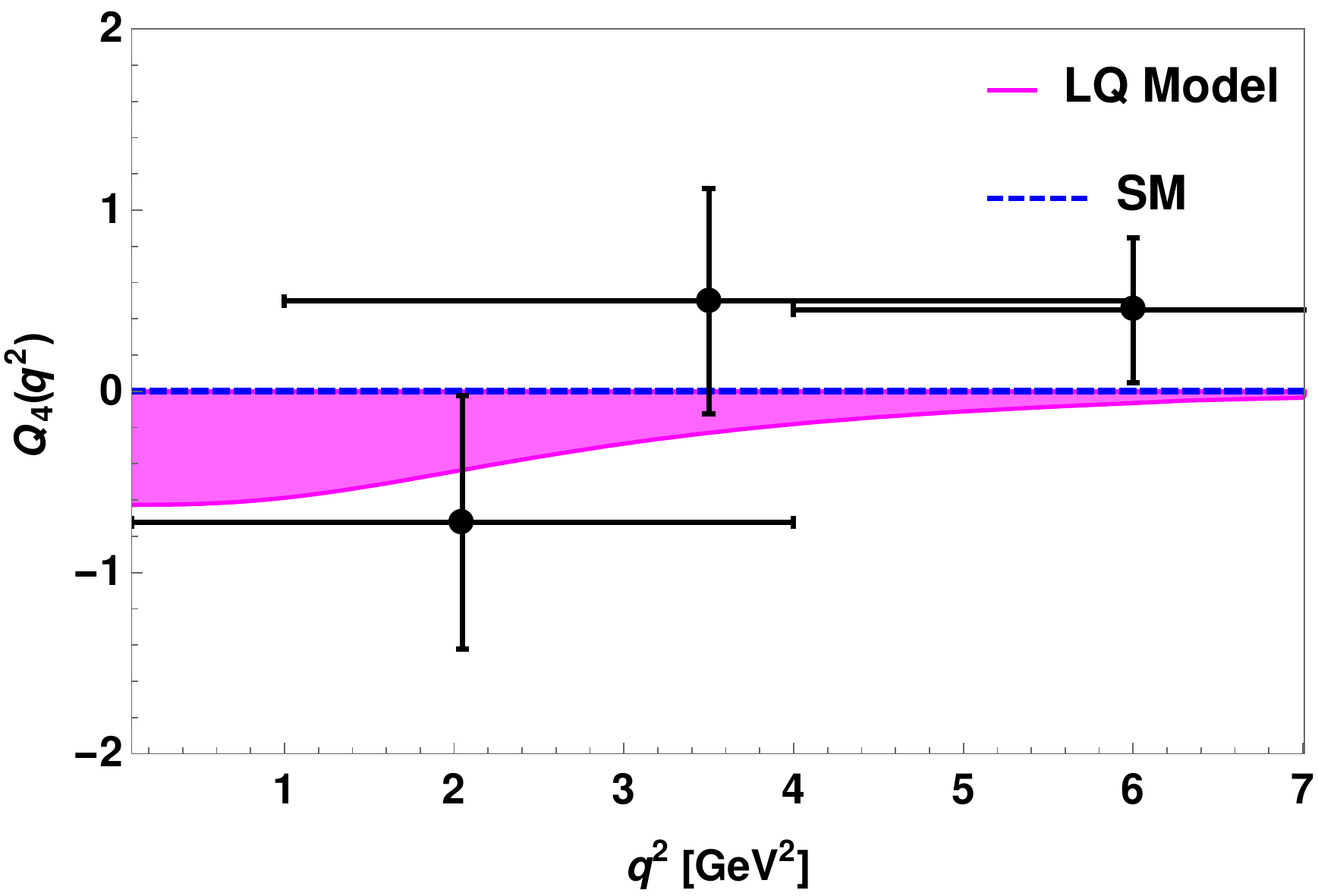}
\quad
\includegraphics[scale=0.45]{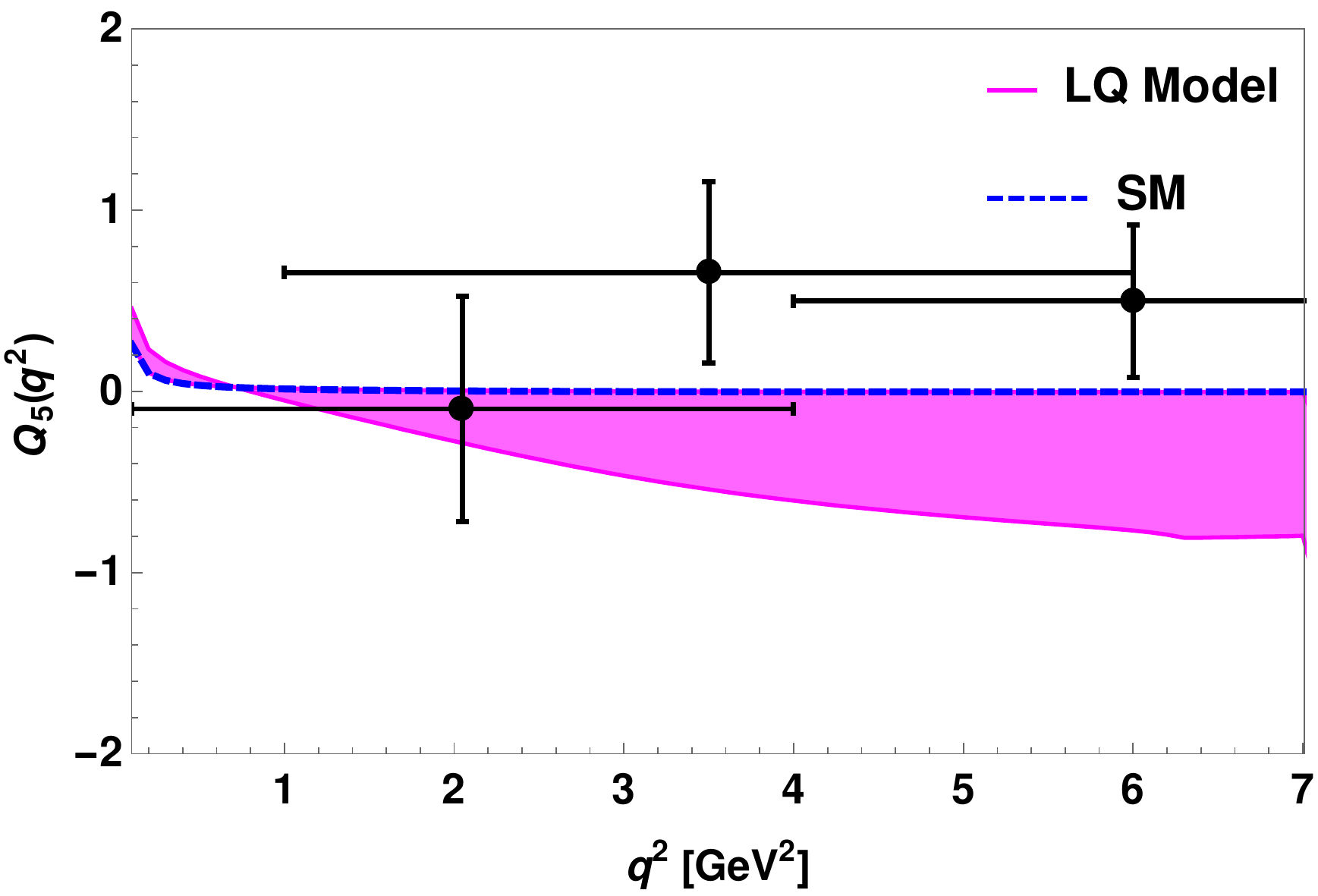}\\
\includegraphics[scale=0.45]{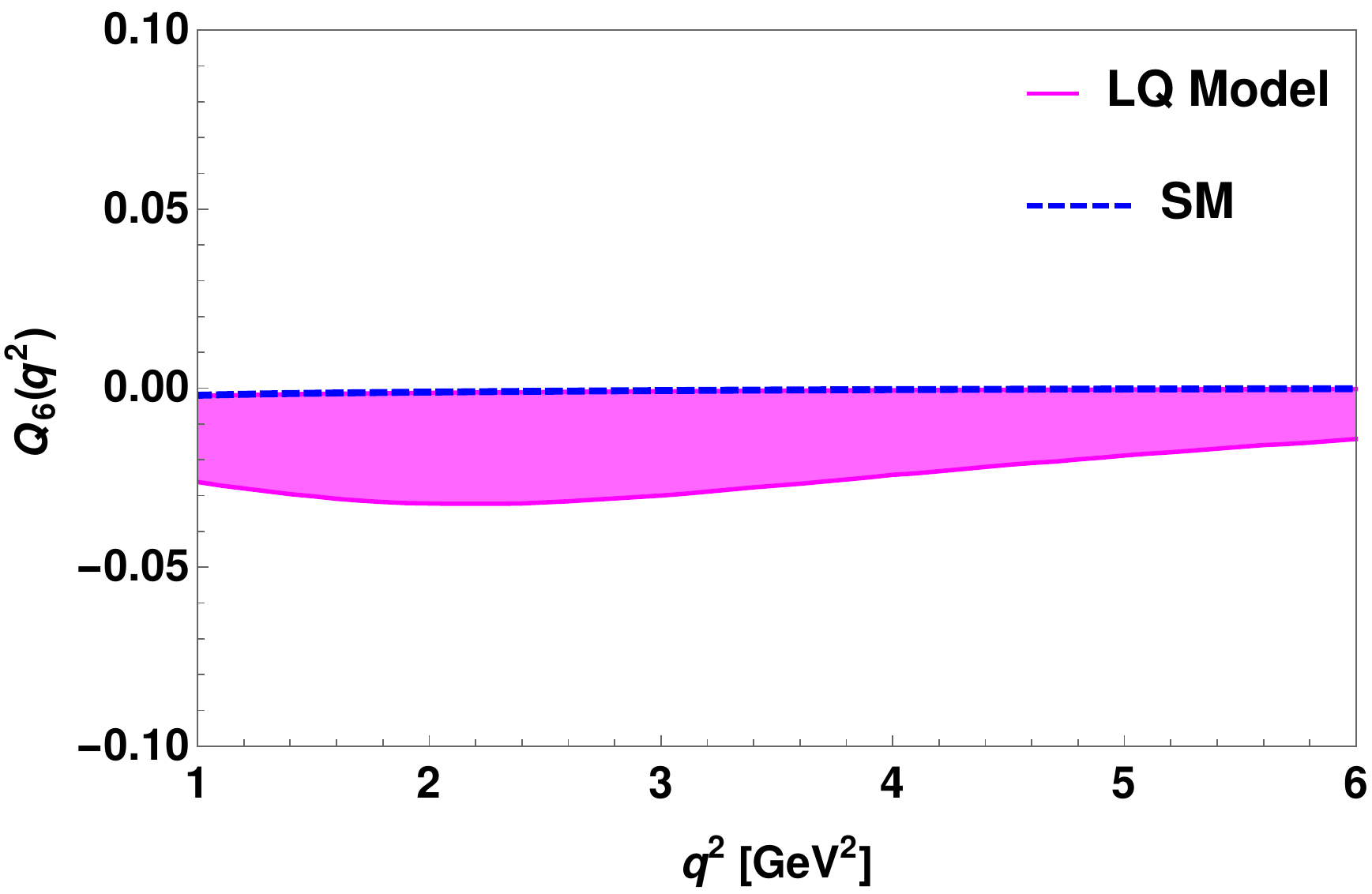}
\quad
\includegraphics[scale=0.45]{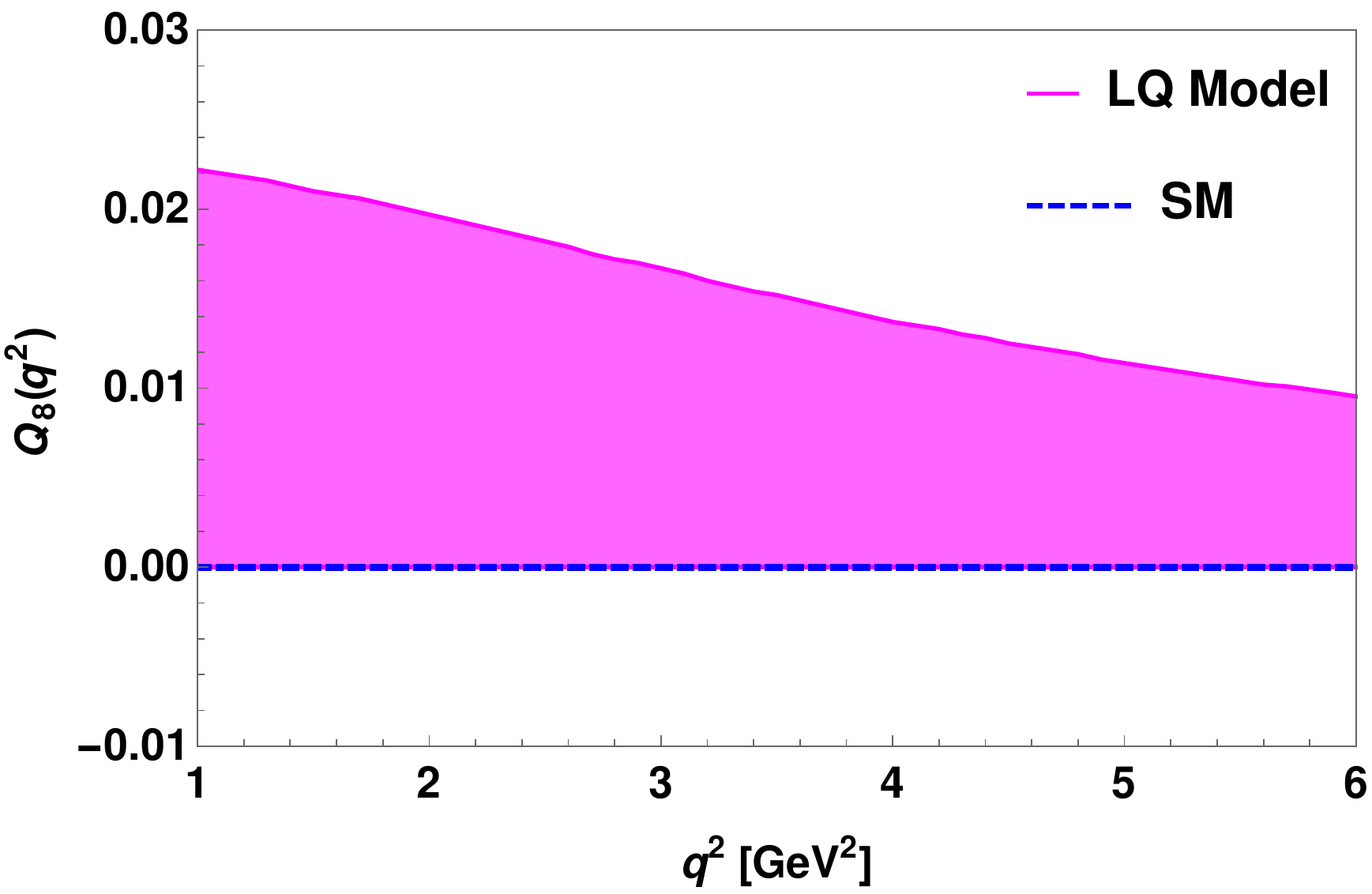}
\caption{The plots in the left panel represent the $Q_{1} (q^2)$ (top), $Q_{4} (q^2)$ (middle) and   $Q_{6} (q^2)$ (bottom)  observables    in the vector LQ model. The $Q_{2} (q^2)$ (top), $Q_{5} (q^2)$ (middle) and   $Q_{8} (q^2)$ (bottom)   plots  are shown in the right panel.}
\end{figure}
\begin{figure}[htb]
\centering
\includegraphics[scale=0.45]{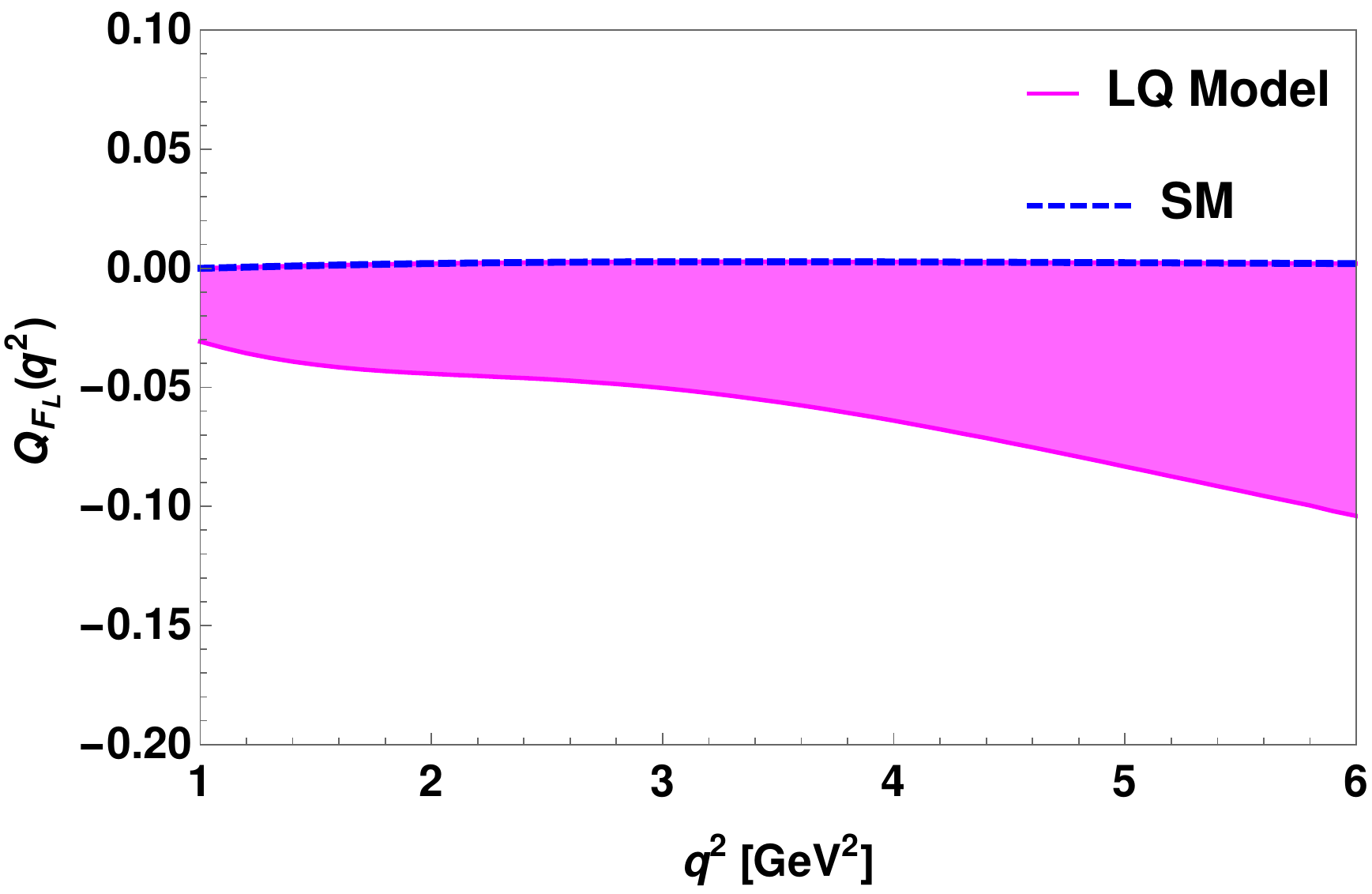}
\quad
\includegraphics[scale=0.45]{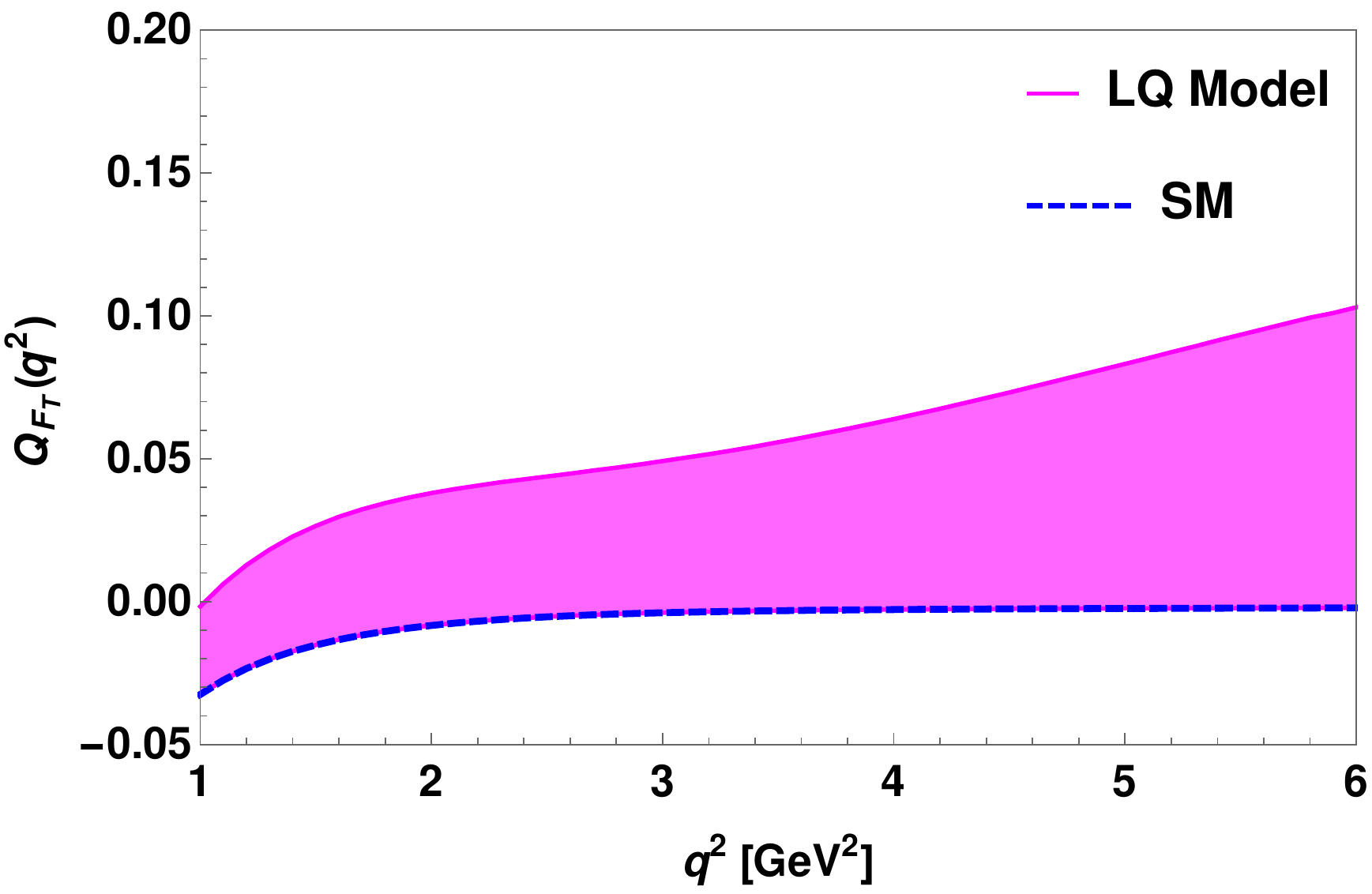}
\caption{The $q^2$ variations of $Q_{F_L} (q^2)$ (left panel) and $Q_{F_T} (q^2)$ (right panel) observables  in the vector LQ model. }
\end{figure}
\begin{figure}[htb]
\centering
\includegraphics[scale=0.45]{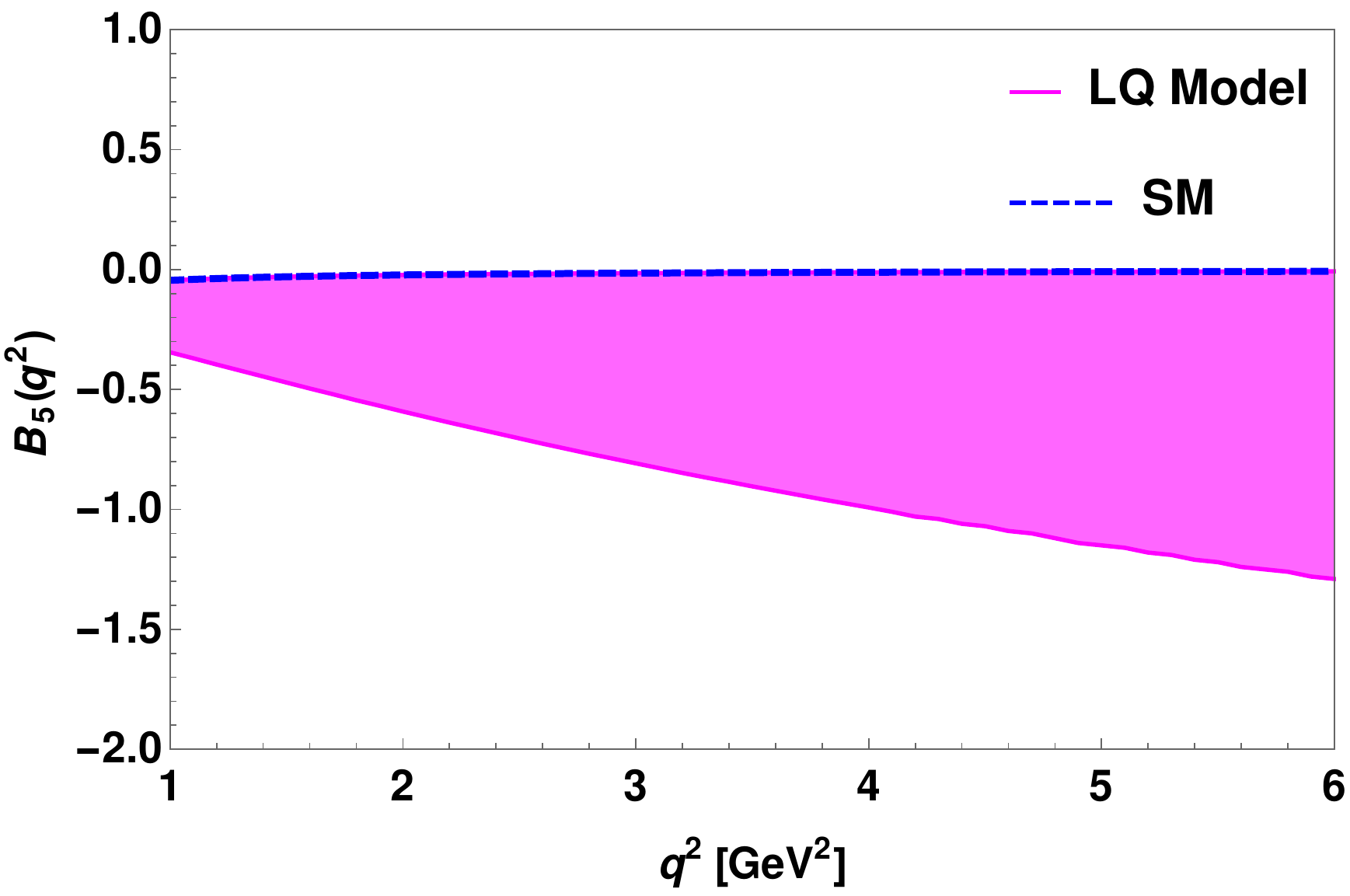}
\quad
\includegraphics[scale=0.45]{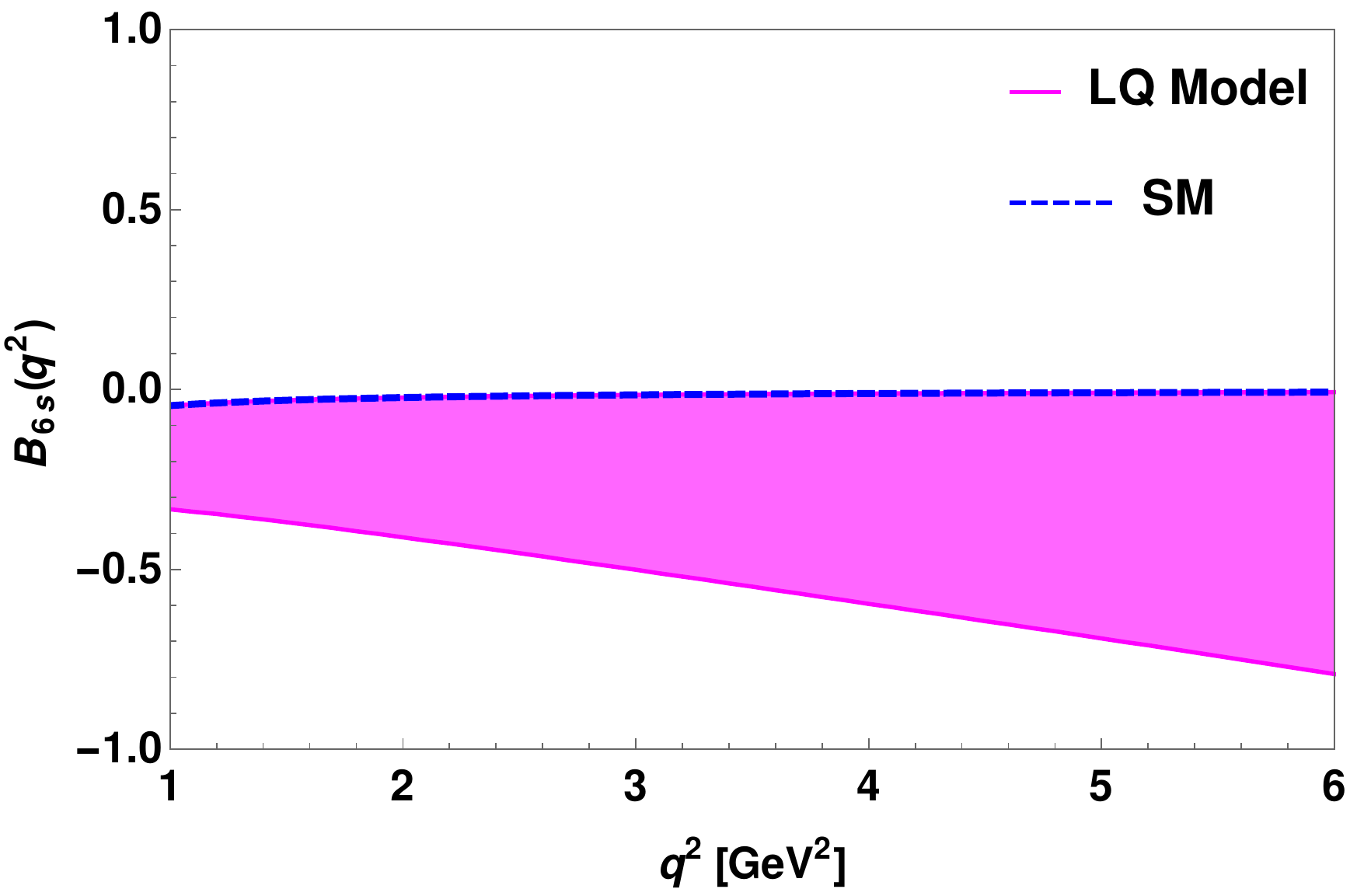}
\caption{The $q^2$ variation of $B_5$ (left panel) and $B_{6s}$ (right panel) observables in the LQ model.}
\end{figure}

\begin{table}[htb]
\begin{center}
\caption{The predicted  values of the $P_i^{l^{(\prime)}}$ observables  in the  low $q^2$ $(q^2 \in [1,6]~{\rm GeV}^2)$ region  for the $\bar B \rightarrow \bar  K^* l^+ l^-$ processes in the SM and the LQ model. }
\begin{tabular}{|c | c | c| }
\hline
 Observables & SM prediction & Values in  LQ model \\
 \hline
 \hline

 $\langle P_1^e \rangle$ & $-0.045 \pm 0.0027$ &  $-0.0448 \to 0.15$ \\
 
 $\langle P_2^e \rangle$ & $0.188 \pm 0.011$& $0.19 \to 0.415$  \\
 
 $\langle P_3^e \rangle$ &$ (-5.43 \pm 0.326) \times 10^{-4}$ & $-0.0143 \to -5.43 \times 10^{-4}$ \\
 
 $\langle P_4^{e \prime} \rangle$ & $0.43 \pm 0.0258$ & $0.43 \to 0.791$ \\
 
 $\langle P_5^{e \prime}  \rangle$ & $-0.226 \pm 0.0136$ & $-0.226 \to 0.682$ \\
 
 $\langle P_6^{e \prime}  \rangle$ & $-0.0734 \pm 0.0044$ & $-0.0734 \to -0.042$ \\
 
 $\langle P_8^{e \prime}  \rangle$ & $0.02678 \pm 0.0016$ & $-0.014 \to 0.027$ \\
 
 \hline
 
 $\langle P_1^\mu \rangle$ & $-0.045 \pm 0.0027$ &  $-0.0449 \to 0.133$ \\
 
 $\langle P_2^\mu \rangle$ & $0.185 \pm 0.011$& $0.185 \to 0.34$  \\
 
 $\langle P_3^\mu \rangle$ & $(-5.41 \pm 0.324) \times 10^{-4}$& $-0.013 \to -5.41 \times 10^{-4}$\\
 
 $\langle P_4^{\mu \prime} \rangle$ & $0.437 \pm 0.026$ & $0.44 \to 0.536$ \\
 
 $\langle P_5^{\mu \prime}  \rangle$ & $-0.2318 \pm 0.014$ &  $-0.231 \to 0.166$\\
 
 $\langle P_6^{\mu \prime}  \rangle$ & $-0.0738 \pm 0.0044$ &  $-0.0676 \to -0.074$\\
 
 $\langle P_8^{\mu \prime}  \rangle$ & $0.02676 \pm 0.001$ &  $1.164 \times 10^{-3} \to 0.027$\\

 \hline
 \end{tabular}
\end{center}
\end{table}
After collecting all possible angular  observables, we now move on for the numerical analysis.  We have taken all the  particle masses and the lifetime of $B$ meson from  \cite{pdg} for the numerical estimation.  We consider  the Wolfenstein parametrization with the values $A =0.811 \pm 0.026$, $\lambda=0.22506\pm 0.00050$,  $\bar \rho = 0.124 ^{+0.019}_{-0.018}$, and $\bar \eta = 0.356 \pm 0.011$ \cite{pdg} for the CKM matrix elements. The QCD form factors for the $\bar B \to \bar K^* l^+ l^-$ processes  in the low $q^2$ region are taken from \cite{kstar-formfactor-1, kstar-formfactor-2}.  Now using the constraints on the   LQ couplings as discussed in section III, we show in Fig. 2, the $q^2$ variation of  the differential  branching ratios of $\bar B \to \bar K^* e^+ e^-$ (left panel) and $\bar B \to \bar K^* \mu^+ \mu^-$ (right panel) processes   in the $V^1(3, 1, 2/3)$ vector LQ  model. In the figures,  the blue dashed lines stand for the SM contributions and the magenta bands are due to the exchange of vector LQ. 
Here the grey bands represent the  theoretical uncertainties, which arise due to the  uncertainties associated with the SM input parameters, such as the  CKM matrix elements \cite{pdg} and the hadronic form factors \cite{kstar-formfactor-1, kstar-formfactor-2}. 
From these
figures, one can see that there is certain difference between the new physics contributions to the  branching fractions  of $\bar B \to \bar K^* e^+ e^-$   and $\bar B \to \bar K^* \mu^+ \mu^-$ processes.    The predicted numerical values of the  branching ratios in the high recoil limit   are presented in Table IV.  In the SM, the forward-backward asymmetry parameters of  the $\bar B \to \bar K^* l^+ l^-$  processes   have negative values in the low-$q^2$ region. However, the contribution of new Wilson coefficients  ($C_{9, 10}^{(\prime)}$ and $C_{S, P}^{(\prime)}$) to the SM due to the exchange of  $(3, 1, 2/3)$ vector LQ 
 may enhance the rate of forward-backward asymmetries and can shift the zero position of these asymmetries.  The plots for the forward-backward asymmetry for the $\bar B \to \bar K^* e^+ e^-$ (left panel) and $\bar B \to \bar K^* \mu^+ \mu^-$ (right panel) processes are presented in Fig. 3 and the corresponding integrated values are given in Table IV. For both $B \to K^* e^+ e^- (\mu^+ \mu^-)$  processes, we found that due to the LQ contributions  the zero-crossing position of forward-backward asymmetry  shifts to the right (i.e., towards high $q^2$ region)  of its SM predicted  value.  The  longitudinal and transverse polarisation components for the  $\bar B \to \bar K^* e^+ e^-$ (left panel) and  $\bar B \to \bar K^* \mu^+ \mu^-$ (right panel) processes both in the SM and in the LQ model   are shown in the Fig. 4 and Fig. 5 respectively. The predicted  values of  $F_L^l(F_T^l)$ asymmetry parameters in the LQ model are given in Table IV. In these observables also we found  some difference between  the SM values and the LQ  contributions.

 
 \begin{table}[htb]
\begin{center}
\caption{The predicted   values of the LFUV observables, ($Q_{F_{L,T}}$,  $Q_i$ and $B_{5, 6s}$)  for  the $B \rightarrow K^* l^+ l^-$ processes in the SM and the LQ model. }
\begin{tabular}{|c | c | c| }
\hline
 Observables & SM prediction & Values in  LQ model \\
 \hline
 \hline

  $\langle Q_{1} \rangle$ & $0$ & $-0.017 \to -0.0001$  \\
 
 $\langle Q_{2} \rangle$ & $-0.003$ & $-0.075 \to -0.005$ \\
 
 $\langle Q_{3} \rangle$ & $2 \times 10^{-6}$ & $2 \times 10^{-6} \to 1.3 \times 10^3$  \\
 
 $\langle Q_{4} \rangle$ & $0.007$  & $-0.255 \to 0.01$  \\
 
 $\langle Q_{5} \rangle$ & $-0.0058$ & $-0.516 \to -0.005$  \\
 
 $\langle Q_{6} \rangle$ &  $-4 \times 10^{-4}$ & $-0.032 \to 5.8 \times 10^{-3}$  \\
 
 $\langle Q_{8} \rangle$ &  $-2 \times 10^{-5}$ & $0 \to 0.0152$  \\ 
 
\hline 

$\langle Q_{F_L} \rangle$ & $0.07$ & $-0.04 \to 0.07$  \\
 
 $\langle Q_{F_T} \rangle$ & $-0.007$ & $-0.007 \to 0.05$  \\ 
 
 \hline

$\langle B_5 \rangle$ & $1.25 \times 10^{-3}$ & $-0.85 \to -1.27 \times 10^{-3}$ \\

$\langle B_{6s} \rangle$ & $-0.027$ & $-0.56 \to -0.027$  \\

\hline
 
\end{tabular}
\end{center}
\end{table}

\vspace*{0.3 truein}

In  Fig. 6, we show the plot for the $R_{K^*}$  observable in the low $q^2$ regime in both the SM and vector  LQ model.  After the $q^2 \sim 1.1~{\rm GeV^2} $ region, noticeable    difference from the SM prediction is found  due to the contribution of the vector LQ. From the figure it can be seen that the measured value of   $R_{K^*}$ in the $q^2 \in [1.1, 6.0]~{\rm GeV}^2$ region can be described in the LQ model.  The predicted values of $R_{K^*}$  in the LQ model for different bins  are presented in Table V.  We found that our predicted results in the vector LQ model are consistent with  the corresponding measured experimental data. Thus, vector LQs could be considered as  potential candidates to explain an possibly   lepton flavour universality violation, should it be observed.

 Fig. 7 shows the  plots for the FFI  observables,  $P_5^{\prime l}$  with respect to  $q^2$ in the large recoil limit. In this figure, the plot for   $P_5^{\prime l}$  for the electron mode is presented in the left panel and the right panel contains the corresponding plot for $\bar B \to \bar K^* \mu^+ \mu^-$ process. 
One can notice that, the LQ model encompasses the SM, but also exhibits potentially larger values of $P_{5}^{ \prime  l}$ observables.
In Table VI,  we have presented the corresponding  numerical results.  In  addition to the  $P_5^{\prime  l}$ observable,  we have also studied  all the FFI observables, $P_{i}^{ (\prime) l}$, where $i=1,2,3,4,6,8$ and the predicted numerical values are listed in Table VI.

The measurement of  $R_{K^*}$  motivated us to look for other LFUV parameters in this process. Belle has recently measured the new LFUV $Q_4$ and $ Q_5$ parameters \cite{Q4-Belle} in the low $q^2$ region, $(1\leq q^2 \leq 6)~{\rm GeV}^2$ with values
\bea
Q_4 = 0.498 \pm 0.527 \pm 0.166, ~~~~~Q_5=0.656 \pm 0.485 \pm 0.103.
\eea
  The $q^2$ variation of $Q_{i}$ parameters in the vector LQ model are presented Fig. 8. In this figure, the left panel contains the plots for  the $Q_1(q^2)$ (top),  $Q_4(q^2)$ (middle) and $Q_6(q^2)$  (bottom) observables and the   $Q_2(q^2)$ (top),  $Q_5(q^2)$ (middle) and $Q_8(q^2)$ (bottom) plots  are given in the right panel.  We observe that the additional contributions due to   LQ has provided large  shift in  some of these observables from their SM values.  
In Fig. 9, we show the plots for $Q_{F_L}$ (left panel) and $Q_{F_T}$ (right panel) observables. We also show the plots for the  $B_5$ (left panel) and $B_{6s}$ (right panel) parameters in Fig. 10.  The numerical values of all these LFUV parameters are given in Table VII.

\section{$K_L \to \mu^\mp e^\pm $ process}

The $V^1_\mu(3,1,2/3)$ vector LQ has also contribution to the lepton flavour violating  $K_{L} \to \mu^\mp e^\pm$ decay process. 
The effective Hamiltonian for $K_{L} \to \mu^- e^+$ LFV decays in the $(3,1,2/3)$ leptoquark model is given by 
\bea
\mathcal{H}_{\rm LQ}&=& C_{LL} \left(\bar{d}\gamma^\mu \left(1-\gamma_5\right)s\right) \left(\bar{\mu}\gamma_\mu \left(1-\gamma_5\right)e\right)+
C_{RR} \left(\bar{d}\gamma^\mu \left(1+\gamma_5\right)s\right) \left(\bar{\mu}\gamma_\mu \left(1+\gamma_5\right)e\right) \nn \\ & + &
C_{LR} \left(\bar{d} \left(1+\gamma_5\right)s\right) \left(\bar{\mu} \left(1-\gamma_5\right)e\right)+
C_{RL} \left(\bar{d} \left(1-\gamma_5\right)s\right) \left(\bar{\mu} \left(1+\gamma_5\right)e\right), ~~
\eea
where  the $C_{LL}, C_{RR}, C_{LR}~{\rm and}~ C_{RL}$ coefficients are given as
\bea \label{KL-Cof}
&&C_{LL}=\frac{(g_L)_{de} (g_L)_{s\mu }^*}{4M_{\rm LQ}^2}, ~~~~~~~~C_{RR}=\frac{(g_R)_{de} (g_R)_{s\mu }^*}{4M_{\rm LQ}^2}, \nn \\ 
&&C_{LR}=\frac{(g_L)_{de} (g_R)_{s\mu }^*}{2M_{\rm LQ}^2}, ~~~~~~~~~C_{RL}=\frac{(g_R)_{de} (g_L)_{s\mu }^*}{2M_{\rm LQ}^2},
\eea
and for  $K_{L} \to \mu^+ e^-$ process
\bea
\mathcal{H}_{\rm LQ}&=& D_{LL} \left(\bar{d}\gamma^\mu \left(1-\gamma_5\right)s\right) \left(\bar{e}\gamma_\mu \left(1-\gamma_5\right) \mu \right)+
D_{RR} \left(\bar{d}\gamma^\mu \left(1+\gamma_5\right)s\right) \left(\bar{e}\gamma_\mu \left(1+\gamma_5\right) \mu \right) \nn \\ & + &
D_{LR} \left(\bar{d} \left(1+\gamma_5\right)s\right) \left(\bar{e} \left(1-\gamma_5\right) \mu \right)+
D_{RL} \left(\bar{d} \left(1-\gamma_5\right)s\right) \left(\bar{e} \left(1+\gamma_5\right) \mu \right), ~~
\eea
where  the $D_{LL}, D_{RR}, D_{LR}~{\rm and}~ D_{RL}$ coefficients are given as
\bea \label{KL-Cof}
&&D_{LL}=\frac{(g_L)_{d\mu} (g_L)_{se }^*}{4M_{\rm LQ}^2}, ~~~~~~~~D_{RR}=\frac{(g_R)_{d\mu} (g_R)_{se}^*}{4M_{\rm LQ}^2}, \nn \\ 
&&D_{LR}=\frac{(g_L)_{d\mu} (g_R)_{se }^*}{2M_{\rm LQ}^2}, ~~~~~~~~~D_{RL}=\frac{(g_R)_{d\mu} (g_L)_{se}^*}{2M_{\rm LQ}^2}.
\eea
The LFV decay processes do not receive any contribution from the SM. In the literature \cite{KL-LQ, KL-LFV}, the LFV decay of kaon has been investigated in the leptoquark and other new physics models.
The  branching ratio of $K_L \to \mu^- e^+$ process in the leptoquark model is given by
\bea \label{KL-BR}
{\rm BR}(K_L \to \mu^- e^+) &=& \tau_{K_L}\frac{f_K^2}{8\pi M_K^3 } \sqrt{\left(M_K^2 -m_\mu^2 - m_e^2\right)^2-4m_\mu^2 m_e^2}  \times \nn \\ && \Bigg [ \Big |( C_{LL} + C_{RR}) (m_e-m_\mu)- (C_{LR} + C_{RL} ) \frac{M_K^2}{m_s+m_d} \Big |^2  \left(M_K^2 - (m_\mu + m_e)^2 \right) \nn \\ && +  \Big | ( C_{LL} - C_{RR}) (m_e + m_\mu)+ (C_{LR} - C_{RL} ) \frac{M_K^2}{m_s+m_d} \Big |^2    \left(M_K^2 - (m_\mu - m_e)^2 \right) \Bigg ].\nn \\
\eea
Similarly, the branching ratio of $K_L \to \mu^+ e^-$ process can be obtained from Eqn. (\ref{KL-BR}) by replacing the new coefficients  $C_{ij} \to D_{ij}$, where ($i,j=L,R$).     The branching ratio of $K_L \to \mu^\mp e^\pm$ process is simply the sum of the branching ratios of  $K_L \to \mu^- e^+$ and $K_L \to \mu^+ e^- $ processes.  For the required LQ couplings, we use the couplings obtained from $K_L \to e^+ e^- (\mu^+ \mu^-)$ process which are given in Table II and III as basis values and assumed that the LQ couplings between different generation of
quark and lepton follow the simple scaling law, i.e., $(g_{L(R)})_{ij} = (m_i /m_j )^{1/2} (g_{L(R)} )_{ii}$ with $j > i$. We have taken this ansatz from the Ref. \cite{ansatz}, which successfully explains the decay width of
radiative LFV $\mu \to e \gamma$ decay.
Now using this ansatz and the particle masses and life time of $K_L$ meson from \cite{pdg}, the predicted branching ratios of $K_{L} \to \mu^\mp e^\pm$ process is given by
\bea
{\rm BR}(K_L \to \mu^\mp e^\pm)&=& (1.78-3.564) \times 10^{-12}.
\eea
The corresponding experimental upper limit on branching ratio of $K_L \to \mu^\mp e^\pm$ process is given as \cite{pdg}
\bea
{\rm BR}(K_L \to \mu^\mp e^\pm) ~ \textless ~4.7 \times 10^{-12}. 
\eea  
  Our predicted branching ratios are within the experimental limit.

\section{Conclusion}
We have investigated  the intriguing  anomalies related  with  the   rare  semileptonic $\bar B \to \bar K^* l^+ l^-$ decay processes in the context of a  $(3,1,2/3)$ vector leptoquark model. We constrain the leptoquark couplings by using  the experimental branching ratios of $B_s \to l^+ l^-$, $K_L \to l^+ l^-$ and $B_s \to \mu^\mp e^\pm$ processes. We then calculated  the branching  ratios, 
forward-backward asymmetries and the lepton polarization asymmetries of these processes.  
We found that there is appreciable difference between the SM and LQ model predictions.
 We have also calculated the form factor independent observables $P_i^{(\prime)}$, where $i=1,..,6,8$   in this model. We observed that vector leptoquark can also explain the $P_{5}^\prime$ anomalies very well.

We then looked into the lepton nonuniversality parameter $R_{K^*}$ of  the $\bar B \to \bar K^* l^+ l^-$ process  in both the $q^2 \in [0.045, 1.1]~{\rm GeV}^2$ and $q^2 \in [1.1, 6.0]~{\rm GeV}^2$ regions and found that the $R_{K^*}$ anomaly could be explained in the vector leptoquark model.   We have also investigated a few other lepton nonuniversality observables in order to verify violation of lepton universality in the $B$ sector. Thus, along with the $R_{K^*}$ observable, we also studied some LNU observables, such as $Q_{F_L}$, $Q_{F_T}$, $Q_{i}$ and  $B_{5, 6s}$  in the vector leptoquark model. We observed that in the presence of a vector leptoquark, all the observables have some differences from their SM results but  in many cases the SM results are within the uncertainties of the LQ model. We have also computed the branching ratio of the lepton flavour violating $K_L \to \mu^\mp e^\pm$  process in the $(3,1,2/3)$ vector leptoquark, which is found to be within the experimental limit.  The observation of  these observables  in the LHCb experiment may  provide indirect hints for the possible existence of leptoquark.

{\bf Acknowledgments}

We would like to thank Science and Engineering Research Board (SERB),
Government of India for financial support through grant No. SB/S2/HEP-017/2013.


\begin{thebibliography}{60}

\bibitem{RKstar-exp}
R. Aaij et al., [LHCb Collaboration], JHEP {\bf 08}, 055 (2017) [arXiv:1705.05802].

\bibitem{RK-exp}
R. Aaij et al., [LHCb Collaboration], Phys. Rev. Lett. \textbf{113}, 151601 (2014) [arXiv:1406.6482].

\bibitem{phi-decayrate}
R. Aaij et al., [LHCb Collaboration], JHEP \textbf{1307}, 084 (2013) [arXiv:1305.2168]. 

 \bibitem{P5p}
R. Aaij et al., [LHCb Collaboration], Phys. Rev. Lett. \textbf{111}, 191801 (2013) [arXiv:1308.1707]. 
  

\bibitem{Kstar-decayrate}
R. Aaij et al., [LHCb Collaboration], JHEP \textbf{1406}, 133 (2014) [arXiv:1403.8044].


\bibitem{isospin-kstar}
C. Langenbruch on behalf of the LHCb collaboration, [arXiv: 1505.04160].

 
\bibitem{RDstar-LHCb}
   R.~Aaij {\it et al.} [LHCb Collaboration],
  Phys.\ Rev.\ Lett.\  {\bf 115},  111803  (2015) 
   Addendum: Phys.\ Rev.\ Lett.\  {\bf 115},  159901 (2015) [arXiv:1506.08614].
   
   
\bibitem{RD-BaBar}
BaBar Collaboration, J. Lees {\it et al.}, Phys. Rev. Lett. \textbf{109}, 101802 (2012)  [arXiv:1205.5442]; 
BaBar Collaboration, J. Lees {\it et al.},  Phys. Rev. D \textbf{88}, 072012 (2013) [arXiv:1303.0571];
Belle Collaboration,   M.~Huschle {\it et al.},
  Phys.  Rev.  D {\bf 92}, 072014 (2015)  [arXiv:1507.03233];
Belle Collaboration, A.~Abdesselam {\it et al.},
  [arXiv:1603.06711].
 
 
 \bibitem{RD-exp}
  Heavy Flavour Averaging Group, 
http://www.slac.stanford.edu/xorg/hfag/semi/winter16/\\winter16\_dtaunu.html.

\bibitem{RK-SM} 
C. Bobeth, G. Hiller, G. Piranishvili,
 JHEP {\bf 12}, 040 (2007) [arXiv:0709.4174]. 
 
\bibitem{RKstar-SM}
B. Capdevila, A. Crivellin, S. Descotes-Genon, J. Matias, and  J. Virto, JHEP {\bf 1801}, 093 (2018) [arXiv:1704.05340]. 

\bibitem{RD-SM}  
  H. Na {\it et al.}, Phys. Rev. D \textbf{92}, 054410 (2015) [arXiv:1505.03925].


  
\bibitem{RDstar-SM}
S.Fajfer, J.F.Kamenik, and I.Nisandzic, Phys. Rev. D \textbf{85}, 094025 (2012) [arXiv:1203.2654];  S.~Fajfer, J.~F.~Kamenik, I.~Nisandzic and J.~Zupan,
  Phys.\ Rev.\ Lett.\  {\bf 109}, 161801 (2012)
  [arXiv:1206.1872].
  

\bibitem{Q4-Belle}
S. Wehle et al. [Belle Collaboration],  Phys.
Rev. Lett. \textbf{118}, 111801 (2017)  [arXiv:1612.05014].

\bibitem{recent-arXiv}

G. D'Amico, M. Nardecchia, P. Panci, F. Sannino, A.  Strumia, R. Torre, and A. Urbano, JHEP {\bf 09}, 010 (2017) [arXiv:1704.05438]; G. Hiller, I. Nisandzic, Phys. Rev. D {\bf 96}, 035003 (2017) [arXiv:1704.05444]; L.-S. Geng, B. Grinstein, S. J$\ddot a$ger, J. M. Camalich, X.-L. Ren, and  R.-X. Shi, Phys. Rev. D {\bf 96}, 093006 (2017) [arXiv:1704.05446]; M. Ciuchini, A. M. Coutinho, M. Fedele, E. Franco, A. Paul, L. Silvestrini, and M. Valli,  Eur. Phys. J. C {\bf 77},  688 (2017)  [arXiv:1704.05447]; A. Celis, J. Fuentes-Martin, A. Vicente, and J. Virto, Phys. Rev. D {\bf 96}, 035026 (2017)  [arXiv:1704.05672]; D. Becirevic, and O. Sumensari, JHEP {\bf 1708}, 104 (2017)  [arXiv:1704.05835]; J. F. Kamenik, Y. Soreq, and J. Zupan, Phys. Rev. D {\bf 97}, 035002 (2018)  [arXiv:1704.06005]; F. Sala, and D. M. Straub, Phys. Lett. B {\bf 774}, 205 (2017)  [arXiv:1704.06188]; S. D. Chiara, A. Fowlie, S. Fraser, C. Marzo, L. Marzola, M. Raidal, and C. Spethmann, Nucl. Phys. B {\bf 923}, 245 (2017) [arXiv:1704.06200]; D. Ghosh, [arXiv:1704.06240]; W. Altmannshofer, P. S. B. Dev, and A. Soni,  Phys. Rev. D {\bf 96}, 095010 (2017)  [arXiv:1704.06659]; A. K. Alok, B. Bhattacharya, A. Datta, D. Kumar, J. Kumar, and  D. London,  Phys. Rev. D {\bf 96}, 095009  (2017)  [arXiv:1704.07397]; F. Bishara, U. Haisch and  P. F. Monni, Phys. Rev. D {\bf 96}, 055002 (2017) [arXiv:1705.03465]; T. Hurth, F. Mahmoudi, D. M. Santos and S. Neshatpour, Phys. Rev. D {\bf 96}, 095034 (2017)  [arXiv:1705.06274];  A. Datta, J. Kumar, J. Liao and D. Marfatia, [arXiv:1705.08423];
 D. Bardhan, P. Byakti and D. Ghosh, Phys. Lett. B {\bf 773}, 505 (2017)  [arXiv:1705.09305]; S. Neshatpour, V.G. Chobanova, T. Hurth, F. Mahmoudi and D. M.  Santos, [arXiv:1705.10730]; S. Matsuzaki, K. Nishiwaki and R. Watanabe, JHEP {\bf 1708}, 145 (2017)  [arXiv:1706.01463]; C.-W. Chiang, X.-G. He, J. Tandean and X.-Bo Yuan, Phys. Rev. D {\bf 96}, 115022 (2017) [arXiv:1706.02696]; J. Kawamura, S. Okawa and Y. Omura, Phys. Rev. D {\bf 96}, 075041 (2017)  [arXiv:1706.04344]; B. Chauhan, B. Kindra and A.  Narang, [arXiv:1706.04598]; S. Khalil, [arXiv:1706.07337]. 

\bibitem{mohanta0}

 S. Sahoo and R. Mohanta, Phys. Rev. D \textbf{93}, 034018 (2016)  [arXiv:1507.02070];

\bibitem{kstar-Q4}
B. Capdevila, S. Descotes-Genon, J. Matias and
J. Virto,  JHEP \textbf{1610}, 075 (2016) 
[arXiv:1605.03156].


\bibitem{GUT}
H. Georgi, and S. L. Glashow, Phys. Rev. Lett. \textbf{32}, 438 (1974); 
 H. Fritzsch and P. Minkowski, Ann. Phys. \textbf{93}, 193
(1975); P. Langacker, Phys. Rep. \textbf{72}, 185 (1981).

\bibitem{Pati}
J. C. Pati, and A. Salam, Phys.
Rev. D \textbf{10}, 275 (1974).

\bibitem{Pati-salam}
J.C. Pati, and A. Salam, Phys. Rev. D \textbf{8}, 1240(1973); Phys. Rev. Lett.
\textbf{31}, 661 (1973); O. Shenkar, Nucl. Phys. B \textbf{206}, 253 (1982); O. Shenkar, Nucl. Phys. B \textbf{204},  375 (1982). 

\bibitem{Composite}
D. B. Kaplan, Nucl. Phys. B \textbf{365}, 259 (1991).

\bibitem{Technicolor}
B. Schrempp, and F. Shrempp, Phys. Lett. B \textbf{153}, 101 (1985); B. Gripaios, JHEP \textbf{1002}, 045
(2010) [arXiv:0910.1789].



\bibitem{mohanta1}
 R. Mohanta, Phys. Rev. D \textbf{89}, 014020 (2014) [arXiv:1310.0713]; S. Sahoo and R. Mohanta, Phys. Rev. D \textbf{91}, 094019 (2015) [arXiv:1501.05193].
 

\bibitem{mohanta2}

 S. Sahoo and R. Mohanta,  New J. Phys. \textbf{18}, 013032 (2016) [arXiv:1509.06248]; Phys. Rev. D \textbf{93}, 114001  (2016) [arXiv:1512.04657]; New J.Phys. \textbf{18}, 093051 (2016) [arXiv:1607.04449];  J.Phys. G \textbf{44},  035001 (2017) [arXiv:1612.02543]; Eur. Phys. J. C {\bf 77}, 344 (2017) [arXiv:1705.02251]; S. Sahoo, R. Mohanta, and A. K. Giri, Phys. Rev. D \textbf{95},  035027  (2017) [arXiv:1609.04367]. 
 
 \bibitem{mohanta3}
 M. Duraisamy, S. Sahoo, and R. Mohanta, Phys. Rev. D \textbf{95},  035022  (2017) [arXiv:1610.00902].
  
 
  \bibitem{davidson}
   S. Davidson, D. C. Bailey and B. A. Campbell, Z. Phys.
C \textbf{61}, 613 (1994) [arXiv:hep-ph/9309310]; I. Dorsner, S. Fajfer, J. F. Kamenik, N. Kosnik, Phys. Lett. B \textbf{682}, 67 (2009) [arXiv:0906.5585]; S. Fajfer, N. Kosnik, Phys. Rev. D \textbf{79}, 017502 (2009) [arXiv:0810.4858]; R. Benbrik, M. Chabab, G. Faisel, [arXiv:1009.3886]; A. V. Povarov, A. D. Smirnov, [arXiv:1010.5707]; J. P Saha, B. Misra and A. Kundu, Phys. Rev. D \textbf{81}, 095011 (2010) [arXiv:1003.1384];  I. Dorsner, J. Drobnak, S. Fajfer, J. F. Kamenik, N. Kosnik, JHEP \textbf{11}, 002 (2011)  [arXiv:1107.5393]; F. S. Queiroz, K. Sinha, A. Strumia,   Phys. Rev. D \textbf{91}, 035006 (2015) [arXiv:1409.6301]; B. Allanach, A. Alves, F. S. Queiroz,   K. Sinha, A. Strumia, Phys. Rev. D \textbf{92}, 055023 (2015) [arXiv:1501.03494];
 Ivo de M. Varzielas, G. Hiller, JHEP \textbf{1506}, 072 (2005)    [arXiv:1503.01084];  
 I. Dorsner, S. Fajfer, A. Greljo, J. F. Kamenik, N. Kosnik, doi:10.1016/j.physrep.2016.06.001, [arXiv:1603.04993];   S. Fajfer, J. F. Kamenik, I. Nisandzic and J. Zupan, Phys. Rev. Lett. \textbf{109}, 161801 (2012) [arXiv:1206.1872]; M. Freytsis, Z. Ligeti and J. T. Ruderman, Phys. Rev. D
\textbf{92}, 054018 (2015) [arXiv:1506.08896]; I. Dorsner, S. Fajfer, J. F. Kamenik and N. Kosnik, Phys. Lett. B \textbf{682}, 67 (2009) [arXiv:0906.5585];   L. Calibbi, A. Crivellin, T. Ota, Phys. Rev. Lett. \textbf{115}, 181801 (2015) [arXiv:1506.02661]; Xin-Q. Li, Ya-D. Yang, X. Zhang,  [arXiv:1605.09308]; B. Dumont, K. Nishiwaki, R. Watanabe,  Phys. Rev. D \textbf{94}, 034001 (2016) [arXiv:1603.05248]; 
M. Bauer and M. Neubert, Phys. Rev. Lett. \textbf{116}, 141802 (2016)  [arXiv:1511.01900]; S. Fajfer, N. Kosnik, doi:10.1016/j.physletb.2016.02.018, [arXiv:1511.06024]; D. Becirevic, S. Fajfer, N. Kosnik, O. Sumensari,  Phys. Rev. D {\bf 94}, 115021 (2016) [arXiv:1608.08501].

\bibitem{kosnik-LQ}
N. Kosnik, Phys. Rev. D \textbf{86}, 055004 (2012), [arXiv:1206.2970].

\bibitem{KL-LQ}
G. Kumar, 	Phys. Rev. D {\bf 94}, 014022 (2016) [arXiv: 1603.00346].

\bibitem{b-s-Hamiltonian}
A. J. Buras and M. Munz, Phys. Rev. D \textbf{52}, 186 (1995); M. Misiak, Nucl. Phys. B \textbf{393}, 23
(1993); ibid. \textbf{439}, 461 (E) (1995).

\bibitem{b-s-Wilson}
Wei-Shu Hou, M. Kohda and F. Xu, Phys. Rev. D \textbf{90}, 013002 (2014) [arXiv:1403.7410].

\bibitem{Buras}
K. De Bruyn, R. Fleischer, R. Knegjens, P. Koppenburg, M. Merk, A. Pellegrino, and N. Tuning, Phys. Rev. Lett. \textbf{109},
041801 (2012); A. J. Buras, R. Fleischer, J. Girrbach and  R. Knegjens, JHEP \textbf{1307}, 77  (2013)  [arXiv:1303.3820].

\bibitem{Bobeth}
C. Bobeth, M. Gorbahn, T. Hermann, M. Misiak, E. Stamou,
and M. Steinhauser, Phys. Rev. Lett. \textbf{112}, 101801 (2014).

\bibitem{e-br}
T. Aaltonen et al. (CDF Collaboration), Phys. Rev. Lett.
\textbf{102}, 201801 (2009).

\bibitem{mu-br}
 LHCb, CMS Collaboration, V. Khachatryan et al., Nature \textbf{522}, 68 (2015) [arXiv:1411.4413].

\bibitem{tau-br}
LHCb Collaboration, LHCb-CONF-2016-011,  https://cds.cern.ch/record/2220757.

\bibitem{KL-2}
G. Buchalla and A. J. Buras, Nucl. Phys. B \textbf{412}, 106 (1994) [arXiv:hep-ph/9308272].

\bibitem{KL-3}
M. Misiak and J. Urban, Phys. Lett. B \textbf{451}, 161 (1999) [arXiv:hep-ph/9901278]; G. Buchalla
and A. J. Buras, Nucl. Phys. B \textbf{548}, 309 (1999) [arXiv:hep-ph/9901288].

\bibitem{KL-1}
G. Isdori and R. Unterdorfer, JHEP \textbf{0401}, 009 (2004) [arXiv:hep-ph/0311084].



\bibitem{pdg}
C. Patrignani  et al. (Particle Data Group), Chin. Phys. C {\bf 40}, 100001 (2016).



\bibitem{kstar-br-1}
C. Bobeth, G. Hiller and G. Piranishivili, JHEP
\textbf{07}, 106 (2008) [arXiv:0805.2525].

\bibitem{kstar-br-2}
U. Egede, T. Hurth, J. Matias, M. Ramon, W. Reece, JHEP
\textbf{10}, 056 (2010) [arXiv:1005.0571].

\bibitem{kstar-br-3}
U. Egede, T. Hurth, J. Matias, M. Ramon, W. Reece, JHEP
\textbf{11}, 032 (2008) [arXiv:0807.2589].

\bibitem{kstar-fl-2}
 F. Kr ̈uger, J. Matias, Phys. Rev. D
\textbf{71}, 094009 (2005) [arXiv:hep-ph/0502060].

\bibitem{kstar-fl-1}
 F. Beaujean, C. Bobeth, D. van Dyk and C. Wacker, JHEP
\textbf{08}, 030 (2012) [arXiv:1205.1838].



\bibitem{kstar-p1}
 J. Matias, F. Mescia, M. Ramon and J. Virto, JHEP
\textbf{04}, 104 (2012) [arXiv:1202.4266].

\bibitem{kstar-p4p}
S. Descotes-Genon, J. Matias, M. Ramon, J. Virto, JHEP
\textbf{01}, 048 (2013) [arXiv:1207.2753].



\bibitem{kstar-formfactor-1}
 M. Beneke, T. Feldmann and D. Seidel, Eur. Phys. J. C  
\textbf{41}, 173 (2005) [arXiv:hep-ph/0412400].

\bibitem{kstar-formfactor-2}
 P. Ball and R. Zwicky, Phys. Rev. D
\textbf{71}, 014029 (2005) [hep-ph/0412079].


\bibitem{KL-LFV}
A. Crivellin, G. D'Ambrosio, M. Hoferichter, L.  C. Tunstall, 	Phys. Rev. D {\bf 93}, 074038 (2016) [arXiv: 1601.00970].

\bibitem{ansatz}

B. Gripaios, M. Nardecchia, S. A. Renner, JHEP {\bf 1505}, 006 (2015) [arXiv:1412.1791]; S.
Davidson, G. Isidori, and S. Uhlig, Phys. Lett. B {\bf 663}, 73 (2008) [arXiv:0711.3376]; M. Redi,
JHEP {\bf 1309}, 060 (2013) [arXiv:1306.1525].

 
\end{thebibliography}
\end{document}